\PassOptionsToPackage{dvipsnames}{xcolor}
\documentclass[acmsmall,nonacm]{acmart}
\pdfoutput=1

\usepackage{booktabs}   %
\usepackage{subcaption} %
\usepackage[dvipsnames]{xcolor}
\usepackage{bcprules}
\usepackage{collcell}
\usepackage{listings}
\usepackage[scaled]{beramono}
\usepackage{lipsum}
\usepackage{mathpartir}

\usepackage{mathtools}
\usepackage{mdframed}
\usepackage{amsmath}
\usepackage{varioref}
\usepackage{cleveref}
\usepackage{scalerel}
\usepackage{xspace}
\usepackage{calc}
\usepackage{array,varwidth}
\usepackage{makecell}
\usepackage{diagbox}
\usepackage{float}
\usepackage{xfrac}
\usepackage{enumitem}
\usepackage[skip=2.5pt]{caption}
\usepackage{slashbox}
\usepackage{xifthen}
\usepackage{tikz}
\usepackage{tikz-cd}
\usetikzlibrary{arrows}
\usepackage{cancel}
\usepackage{bm}
\usepackage[export]{adjustbox}
\usepackage{pgfplots}
\usepgfplotslibrary{fillbetween}
\usepackage{wrapfig}

\usetikzlibrary{shadows}
\usetikzlibrary{shadows.blur}
\usetikzlibrary{decorations.pathreplacing} %
\tikzcdset{arrow style=tikz,
           diagrams={>=stealth}
         }    %
\tikzcdset{crossing over clearance=.75ex}
\tikzset{hardd/.style={teal,thick}} %
\tikzset{softd/.style={teal, dashed,thick}} %
\tikzset{datad/.style={black,densely dotted}} %
\tikzset{alld/.style={black,thick}} %
\tikzset{expnode/.style={
    asymmetrical rectangle,rounded corners,draw,fill=white,inner sep=2.5pt,
    font=\footnotesize, %
  }} %
\tikzset{blknode/.style={
    asymmetrical rectangle,sharp corners,draw,violet,fill=violet!10,
    font=\footnotesize %
  }} %
\tikzset{dummynode/.style={fill=none,draw=none,asymmetrical rectangle} %
} %
\tikzset{nmute/.style={opacity=.4,blur shadow={shadow opacity=.4}}} %
\tikzset{expn/.style={
    circle,fill=white,draw,minimum size=5pt,inner sep=0pt,outer sep=0pt,
    font=\footnotesize, %
    label={[font=\footnotesize]left:#1}
  }} %
\tikzset{expn/.default={}}
\tikzset{mexpn/.style={opacity=.4,
    circle,fill=white,draw,minimum size=5pt,inner sep=0pt,outer sep=0pt,
    font=\footnotesize, %
    label={[font=\footnotesize,opacity=.4]left:#1}
  }} %
\tikzset{mexpn/.default={}}
\tikzset{rmexpn/.style={opacity=.4,
    circle,fill=white,draw,minimum size=5pt,inner sep=0pt,outer sep=0pt,
    font=\footnotesize, %
    label={[font=\footnotesize,opacity=.4]right:#1}
  }} %
\tikzset{rmexpn/.default={}}
\tikzset{mexpndummy/.style={opacity=.4,
    circle,fill=white,minimum size=5pt,inner sep=0pt,outer sep=0pt,
    label={[font=\footnotesize,opacity=.4]left:#1}
  }} %
\tikzset{mexpndummy/.default={}}

\makeatletter %
\def\arcr{\@arraycr}
\makeatother

\newcommand{\showDOI}[1]{\unskip}

\newtheorem{innercustomgeneric}{\customgenericname}
\providecommand{\customgenericname}{}
\newcommand{\newcustomtheorem}[2]{%
  \newenvironment{#1}[1]
  {%
   \renewcommand\customgenericname{#2}%
   \renewcommand\theinnercustomgeneric{##1}%
   \innercustomgeneric
  }
  {\endinnercustomgeneric}
}
\newcustomtheorem{re-definition}{Definition}
\newcustomtheorem{re-lemma}{Lemma}

\newcommand{\Specsharp}{%
	{\settoheight{\dimen0}{C}Spec\kern-.05em \resizebox{!}{\dimen0}{\raisebox{\depth}{\#}}}}
\newcommand{\Csharp}{%
	{\settoheight{\dimen0}{C}C\kern-.05em \resizebox{!}{\dimen0}{\raisebox{\depth}{\#}}}}

\definecolor{blue-violet}{rgb}{0.54, 0.17, 0.89}
\definecolor{depmap}{HTML}{00007B}
\definecolor{dark-cyan}{HTML}{135579}
\definecolor{magenta}{HTML}{a8264f}

\lstdefinelanguage{Neutral}%
{morekeywords={abstract,%
  case,catch,char,class,%
  def,else,extends,final,finally,for,%
  if,import,implicit,%
  match,module,%
  new,null,%
  object,override,%
  package,private,protected,public,%
  for,public,return,super,%
  this,trait,try,type,%
  val,var,%
  with,while,%
  yield,%
  let,end,%
	in,fun,alloc,inc%
  },%
  mathescape=true,%
  sensitive,%
  keywordstyle={\color{black}\bf\ttfamily},%
  commentstyle=\color{OliveGreen},%
  escapebegin=\color{OliveGreen},
  morecomment=[l]//,%
  morecomment=[s]{/*}{*/},%
  morecomment=[s][\color{darkgray}]{@}{\ },%
  morestring=[b]",%
  morestring=[b]',%
  showstringspaces=false%
}[keywords,comments,strings]%

\lstdefinelanguage{OOPSLA21}%
{morekeywords={abstract,%
  case,catch,char,class,%
  def,else,extends,final,finally,for,%
  if,import,implicit,%
  match,module,%
  new,null,%
  object,override,%
  package,private,protected,public,%
  for,public,return,super,%
  this,throw,trait,try,type,%
  val,var,%
  with,while,%
  yield,%
  let,end,%
	in,fun,alloc,inc%
  },%
  mathescape=true,%
  sensitive,%
  keywordstyle={\color{magenta}\bf\ttfamily},%
  commentstyle=\color{magenta},%
  escapebegin=\color{magenta},
  morecomment=[l]//,%
  morecomment=[s]{/*}{*/},%
  morecomment=[s][\color{magenta}]{@}{\ },%
  morestring=[b]",%
  morestring=[b]',%
  showstringspaces=false%
}[keywords,comments,strings]%

\lstdefinelanguage{PolyRT}%
{morekeywords={abstract,fn,%
  case,catch,char,class,%
  def,else,extends,final,finally,for,%
  if,import,implicit,then,%
  match,module,%
  new,null,%
  object,override,%
  package,private,protected,public,%
  for,public,return,super,%
  this,throw,trait,try,type,%
  val,var,%
  with,while,%
  yield,%
  let,end,%
	in,fun,alloc,%
  synchronized,%
  map,%
  given,using,%
  extension,
  },%
  morekeywords=[2]{@kill},%
  mathescape=true,%
  sensitive,%
  keywordstyle={\color{dark-cyan}\bf\ttfamily},%
  keywordstyle=[2]{\color{magenta}\ttfamily},%
  commentstyle=\color{dark-cyan},%
  stringstyle=\color{darkgray},%
  escapebegin=\color{dark-cyan},%
  morecomment=[l]//,%
  morecomment=[s]{/*}{*/},%
  morestring=[b]",%
  morestring=[b]',%
  showstringspaces=false%
}[keywords,comments,strings]%

\lstdefinestyle{impl}{basicstyle=\scriptsize\selectfont\ttfamily,numbers=left,xleftmargin=2em, escapechar=|}

\newcommand{\trackset}[1]{^{\texttt{\{#1\}}}}
\newcommand{\track}[0]{^{\varnothing}}
\newcommand{\trackvar}[1]{^{\texttt{#1}}}
\newcommand{\trackfresh}{^{\texttt{\ensuremath\vardiamondsuit}}}

\newcommand{\ty}[2][]{\ensuremath{\ifthenelse{\isempty{#1}}{#2}{#2^{\,#1}}}}

\newcommand{\ts}[1][]{\ensuremath{\ifthenelse{\isempty{#1}}{\,\vdash\,}{\,\vdash^{\,#1}\,}}}

\newcommand{\cx}[2][]{\ensuremath{\ifthenelse{\isempty{#1}}{#2}{#2^{\,#1}}}}

\newcommand{\qbot}{\ensuremath{\varnothing}}
\newcommand{\qfresh}{\ensuremath{\vardiamondsuit}}

\newsavebox{\SMALLSTAR}
\savebox{\SMALLSTAR}{\(\raisebox{.25ex}{\(\qfresh\)}\)}

\newsavebox{\OVRLP}
\savebox{\OVRLP}{$\raisebox{.37ex}[0pt][0pt]{$\mathrlap{\hspace{.415ex}\scaleobj{.5}{\vardiamondsuit}}$}\cap$}

\newcommand{\qsat}[1]{\ensuremath{#1\mathord{*}}}

\newcommand{\hole}[1]{\ensuremath{[\,#1\,]}}
\newcommand{\CX}[3][black]{\ensuremath{{\color{#1}#2\ifthenelse{\isempty{#3}}{}{\hole{{\color{black}#3}}}}}}

\newcommand{\ie}{{\em i.e.}\xspace}

\newcommand{\Fsub}{\ensuremath{\mathsf{F}_{<:}}\xspace}

\makeatletter
\newcommand{\artifacturl}[1]{
  \if@ACM@anonymous
    Link to repository removed for double-blind review.
  \else
    \url{#1}
  \fi
}
\makeatother

\colorlet{eff}{magenta}
\newcommand{\FX}[1]{\ensuremath{{\color{eff}#1}}}
\newcommand{\EPS}[1][]{\ifthenelse{\isempty{#1}}{\FX{\bm{\varepsilon}}}{\FX{\bm{\varepsilon_{#1}}}}}
\newcommand{\EPSS}[1][]{\ifthenelse{\isempty{#1}}{\FX{\qsat{\bm{\varepsilon}}}}{\FX{\qsat{\bm{\varepsilon_{#1}}}}}}
\newcommand{\EPSPR}[1][]{\ifthenelse{\isempty{#1}}{\FX{\bm{\varepsilon'}}}{\FX{\bm{\varepsilon'_{#1}}}}}
\newcommand{\EPSSPR}[1][]{\ifthenelse{\isempty{#1}}{\FX{\qsat{\bm{\varepsilon'}}}}{\FX{\qsat{\bm{\varepsilon'_{#1}}}}}}

\colorlet{meff}{Gray}
\newcommand{\M}{\mathnormal{m}}
\newcommand{\MFX}[1]{\ensuremath{{\color{meff}#1}}}
\newcommand{\MV}[1][]{\ifthenelse{\isempty{#1}}{\MFX{\bm{\M}}}{\MFX{\bm{\M_{#1}}}}}

\colorlet{mute}{teal}

\colorlet{lightred}{red!30}

\newcommand{\KILL}{\ensuremath{\mathscr{k}}}

\newcommand{\JUSTK}[1]{\FX{\{\KILL: #1\}}}

\lstdefinelanguage{DOT}%
{morekeywords={val,new},%
  sensitive,%
  morecomment=[l]//,%
  morecomment=[s]{/*}{*/},%
  morestring=[b]",%
  morestring=[b]',%
  showstringspaces=false%
}[keywords,comments,strings]%

\newlength{\trulemargin}
\newlength{\trulewidth}
\newlength{\srulewidth}
\newenvironment{trules}{$\vspace{0.5em}\ba{p{\trulemargin}@{~}p{\trulewidth}@{~}p{\trulemargin}}}{\ea$}
\newenvironment{srules}{$\vspace{0.5em}\ba{p{\trulemargin}@{~}p{\srulewidth}}}{\ea$}

\newcommand{\ba}{\begin{array}}
\newcommand{\ea}{\end{array}}

\newcommand{\ei}{\end{array}}
\newcommand{\bcases}{\left\{\begin{array}{ll}}
\newcommand{\ecases}{\end{array}\right.}

\newcommand{\eg}{{\em e.g.}\xspace}

\setcopyright{cc}
\setcctype{by}
\acmDOI{10.1145/3808323}
\acmYear{2026}
\acmJournal{PACMPL}
\acmVolume{10}
\acmNumber{PLDI}
\acmArticle{245}
\acmMonth{6}
\acmSubmissionID{pldi26main-p532-p}
\received{2025-11-14}
\received[accepted]{2026-04-03}

\makeatletter%
\begin{document}

\title{Typestate via Revocable Capabilities}

\author{Songlin Jia}
\orcid{0009-0008-2526-0438}
\affiliation{%
  \institution{Purdue University}
  \city{West Lafayette}
  \country{USA}
}
\email{jia137@purdue.edu}

\author{Craig Liu}
\orcid{0009-0003-4113-1306}
\affiliation{%
  \institution{Purdue University}
  \city{West Lafayette}
  \country{USA}
}
\email{liu3477@purdue.edu}

\author{Siyuan He}
\orcid{0009-0002-7130-5592}
\affiliation{%
  \institution{Purdue University}
  \city{West Lafayette}
  \country{USA}
}
\email{he662@purdue.edu}

\author{Haotian Deng}
\orcid{0009-0002-7096-2646}
\affiliation{%
  \institution{Purdue University}
  \city{West Lafayette}
  \country{USA}
}
\email{deng254@purdue.edu}

\author{Yuyan Bao}
\orcid{0000-0002-3832-3134}
\affiliation{%
  \institution{Augusta University}
  \city{Augusta}
  \country{USA}
}
\email{yubao@augusta.edu}

\author{Tiark Rompf}
\orcid{0000-0002-2068-3238}
\affiliation{%
  \institution{Purdue University}
  \city{West Lafayette}
  \country{USA}
}
\email{tiark@purdue.edu}

\begin{abstract}
Managing stateful resources safely and expressively is a longstanding challenge
in programming languages, especially in the presence of
aliasing. For example, scope-based constructs like Java's \texttt{synchronized} blocks offer
ease of reasoning, but they restrict expressiveness and parallelism. 
Conversely,
imperative, flow-sensitive approaches enable fine-grained control, but they require
sophisticated typestate analyses and often burden programmers with explicit
state tracking.

In this work, we present a novel approach that unifies
the ease of scoped reasoning with the expressiveness of imperative typestate management.
Our design extends traditional flow-insensitive capability mechanisms to a flow-sensitive setting.
In particular, we decouple capability
lifetimes from lexical scopes, allowing functions to receive, revoke, or return
capabilities in a flow-sensitive manner,
building on existing mechanisms for the safety and ergonomics of
scoped capability programming.

We implement our approach as an extension to the Scala 3 compiler,
leveraging path-dependent types and implicit resolution to enable concise,
statically safe, and expressive typestate programming.
Our prototype generically supports 
a wide range of patterns, including file operations, advanced locking
protocols, DOM construction, and session types,
showing that
expressive and safe typestate management can be achieved with minimal extensions
to an existing language with capability support. \end{abstract}



\maketitle

\lstMakeShortInline[keywordstyle=,%
                    flexiblecolumns=false,%
                    language=PolyRT,
                    basewidth={0.56em, 0.52em},%
                    mathescape=true,%
                    basicstyle=\footnotesize\ttfamily]@

\section{Introduction} \label{sec:intro}

Programs often perform not only pure computations, but also interact with
external environments, observing and mutating \emph{state}.
Typical examples include file I/O, remote procedure calls, and thread
synchronization.
Programming languages support a variety of mechanisms for managing state,
balancing ease of reasoning with the expressiveness of complex patterns.

Consider the database transaction illustrated in \Cref{fig:capalock}.
It first locates a @row@ within a @table@ and then computes a result
using that @row@. To avoid interference, concurrent access to the @table@ and the @row@
must be prevented. Languages such as Java provide
scope-based constructs like @synchronized@, which
automatically acquire and release locks upon entering and exiting a scope.

While scoped constructs relieve users from manually managing the state
of locks, they lack expressiveness for fine-grained control.
As shown in \Cref{fig:flowlock1}, manual, flow-sensitive management of locks
allows the lock for @table@ to be released immediately after acquiring the
lock for @row@, thereby enabling improved parallelism.
In contrast, nested @synchronized@ blocks impose \emph{last-in-first-out}
(LIFO) lifetimes, forcing @table@ to be locked longer than @row@ and
precluding such optimizations.

\begin{figure}[t]
\begin{subcaptionblock}{.49\linewidth}
\begin{tikzpicture}[x=1cm,y=0.6cm]
\node[anchor=north west, text width=6cm, align=left] at (0,7) {
\begin{lstlisting}
synchronized (table) {
  var row = table.locateRow(...);
  synchronized (row) {
    return computeOnRow(row);
  }  // unlock row
}    // unlock table
\end{lstlisting}
};

\draw[thick,blue] (5.8,6.5) -- (5.8,2.8);
\node[blue, right] at (5.8,6.5) {\tiny\texttt{table}};
\node[blue] at (5.8,5.6) {$\bullet$};

\draw[thick,red] (6.0,5.2) -- (6.0,3.4);
\node[red, right] at (6.0,5.2) {\tiny\texttt{row}};
\node[red] at (6.0,4.4) {$\bullet$};

\end{tikzpicture}
\vspace{-1ex}
\caption{Written using scope-based \lstinline@synchronized@ blocks.
Lock lifetimes are managed automatically, whereas
{\color{blue}\scriptsize\texttt{table}} has to be locked longer than
{\color{red}\scriptsize\texttt{row}}.}
\label{fig:capalock}
\end{subcaptionblock} \hfill
\begin{subcaptionblock}{.49\linewidth}
\begin{tikzpicture}[x=1cm,y=0.6cm]
\node[anchor=north west, text width=6cm, align=left] at (0,7) {
\begin{lstlisting}
table.lock()
val row = table.locateRow(...)
table.lockRow(row)
table.unlock()
val result = computeOnRow(row)
row.unlock()
\end{lstlisting}
};

\draw[thick,blue] (5.8,6.5) -- (5.8,4);
\node[blue, right] at (5.8,6.5) {\tiny\texttt{table}};
\node[blue] at (5.8,5.6) {$\bullet$};

\draw[thick,red] (6.0,5.2) -- (6.0,2.8);
\node[red, right] at (6.0,5.2) {\tiny\texttt{row}};
\node[red] at (6.0,3.6) {$\bullet$};

\end{tikzpicture}
\vspace{-1ex}
\caption{Written in imperative style,
{\color{blue}\scriptsize\texttt{table}} can be unlocked once
{\color{red}\scriptsize\texttt{row}} is locked, enabling improved parallelism,
at the cost of being explicit about lock lifetimes.}
\label{fig:flowlock1}
\end{subcaptionblock}
\caption{A database transaction expressed in two different styles, where we
first {\color{blue}locate a row} in the table and
then {\color{red}compute a result} on the row.
{\color{blue}Table-level} and
{\color{red}row-level}
locks are needed for safe concurrency.
In this work, we combine the ease of scoped reasoning (a) with the
expressiveness of imperative code (b).  } \label{fig:firstlocks}
\end{figure}

Nevertheless, neither approach statically enforces
that locks are acquired before invoking functions requiring them:
programmers may omit @synchronized@ blocks or lock objects entirely, and such
code would remain type-correct yet unsafe in concurrent contexts.
In scope-based programming, 
recent work has introduced mechanisms for providing stronger static guarantees 
\cite{DBLP:conf/oopsla/OsvaldEWAR16,DBLP:conf/ecoop/XhebrajB0R22,
DBLP:conf/scala/OderskyBBLL21,DBLP:journals/pacmpl/WeiBJBR24,
DBLP:journals/toplas/BoruchGruszeckiOLLB23}.
Among these, \emph{capability-based} designs emerge to bridge ease of use and safety, such that a resource handle carries not only the resource but also a capability to operate on it.
Functions like @computeOnRow@ can require a lock as a \emph{capability}
argument, accessible only within @synchronized@:
\begin{lstlisting}
synchronized (row) { lock =>      // given lock
  return computeOnRow(row)(lock)  // using lock (can be inferred in Scala)
}                                 // lock becomes inaccessible
\end{lstlisting}
Languages such as Scala further advance this paradigm through \emph{implicit argument
resolution} \cite{DBLP:journals/pacmpl/OderskyBLBMS18},
which allows 
capabilities to be supplied automatically when they are in scope.

Establishing static safety guarantees for imperative code, by contrast,
necessitates sophisticated \emph{typestate analysis} \cite{DBLP:journals/tse/StromY86}
or \emph{session types} \cite{DBLP:journals/csur/HuttelLVCCDMPRT16}.
Without syntactically scoped lifetimes,
the type system must precisely track the state of {each} lock, whether
locked or unlocked, at {every} program point.
Methods and functions then need to specify both the required state of their
arguments and the state transitions they induce:
\begin{lstlisting}
table.lockRow(row)               // table: Locked required! row: Unlocked to Locked
table.unlock()                   // table: Locked to Unlocked
val result = computeOnRow(row)   // row: Locked required!
row.unlock()                     // row: Locked to Unlocked
\end{lstlisting}
In addition, conventional type systems are designed to track \emph{invariants}
instead of \emph{transitions}.
Specialized mechanisms are thus required, and are further complicated by
sharing and aliasing.

\vspace{-2pt}
\paragraph{This Work.}
We present a solution for flow-sensitive typestate tracking that
builds on existing flow-insensitive capability mechanisms, enabling
static safety reasoning for expressive imperative code
through minimal extensions.
Our approach decouples capability lifetimes from lexical scopes,
and allows any function to provide or
revoke capabilities in a flow-sensitive manner, thereby supporting 
effect and state transition tracking 
independently of any specific typestate discipline.

We have implemented our approach as a prototype by extending the Scala 3
compiler. Capabilities are encoded using path-dependent types
\cite{DBLP:conf/oopsla/RompfA16}, enabling precise association between objects
and their states. With our extensions, Scala's implicit resolution mechanism
facilitates the following three interactions between functions and
capabilities:
\begin{itemize}[leftmargin=3ex,topsep=2pt]
\item \textbf{Receiving:}
  Functions with an \emph{implicit argument arrow} (@?=>@, existing Scala
  feature \cite{DBLP:journals/pacmpl/OderskyBLBMS18}) can receive
  capabilities from the calling context without explicit passing.
\item \textbf{Revoking:}
  Functions with a \emph{destructive arrow} (@=!>@) can revoke capabilities
  (\Cref{subsec:killeff}), ensuring they cannot be subsequently accessed
  (inspired by linear types \cite{DBLP:conf/ifip2/Wadler90}).
\item \textbf{Returning:}
  Functions with an \emph{implicit result arrow} (@?<=@) can return capabilities
  (\Cref{subsec:implicitintro}), making them available for implicit resolution
  at the call site of the function
  (new in this work).
\end{itemize}
Combining them, a composite arrow  (@?=!>?@) expresses state transitions.

Our revocation mechanism relies on a destructive effect system
over \emph{descriptive alias tracking}---capturing checking
\cite{DBLP:journals/toplas/BoruchGruszeckiOLLB23,DBLP:journals/pacmpl/XuBO24,DBLP:journals/pacmpl/XuBPO25}
for the implementation, and reachability types
\cite{DBLP:journals/pacmpl/WeiBJBR24,DBLP:journals/pacmpl/BaoWBJHR21}
for the formal aspects, drawing on the effect system of
\citet{DBLP:journals/corr/abs-2510-08939}.
Returning capabilities is implemented via a type-directed ANF
transformation \cite{DBLP:conf/icfp/RompfMO09}.
Focused on implementation and empirical validation in this work, 
we leave a complete formal account combining
dependent object types~\cite{DBLP:conf/oopsla/RompfA16,DBLP:conf/birthday/AminGORS16}
with implicit resolution, ANF transformation, and destructive effects as
future work.

Our prototype supports a generic typestate system capable of expressing 
diverse effects and state transitions, including file operations
(\Cref{sec:stepbystep}), hand-over-hand locks
(\Cref{fig:flowlock1}, \Cref{sec:lockagain}), stateful DOM tree construction
(\Cref{sec:domtree}), and communication protocols
\cite{DBLP:conf/icfp/JespersenML15,DBLP:conf/haskell/PucellaT08}
(\Cref{sec:sessiontypes}).  Building on the
capability-based programming paradigm that accommodates flexible sharing and
aliasing, our approach enables concise, readable code without extensive
annotations.

The remainder of this paper is structured as follows:
\begin{itemize}[leftmargin=3ex]
\item \Cref{sec:stepbystep} informally introduces our approach 
step-by-step, through an example showing capability-based programming with files.
\item \Cref{sec:casestudy} presents additional case studies, demonstrating our
system on realistic typestate programming scenarios, including locks, DOM trees,
and interprocess communication.
\item \Cref{sec:formalimpl} discusses implementation details, intended safety properties, and limitations.
\end{itemize}
Related work is discussed in \Cref{sec:related}, and we conclude in
\Cref{sec:conc}.
We provide additional implementation details in \Cref{appendix:case-study},
and our artifact can be found on
Zenodo \cite{artifact}.%
\section{From Capabilities to Typestate, Step by Step} \label{sec:stepbystep}

In this section, we informally present our key ideas. We
begin by reviewing a scope-based approach to programming with files
in \Cref{subsec:scopedcaps}.
Next, we discuss how to move toward typestate analysis in
\Cref{sec:naive} and observe its limitations.
Finally, %
we elaborate our three key mechanisms towards safe, expressive, and
ergonomic flow-sensitive reasoning in the remainder of this section.

Starting here, we present code in Scala syntax.  Our implementation relies on a
combination of language features.
Scala, as a widely used language, happens to support this particular
combination, making it a natural platform for our experiment.
Nonetheless, we believe the programming patterns discussed
here are language-independent.
Mechanisms serving the same goals as implicits appear in other languages in
different forms, and alias tracking/controlling mechanisms have also been
explored under various contexts.
Thus, the principles explored here could inform future designs once similar
features become available elsewhere.

\subsection{Preliminary: Programming with Scoped Capabilities}
\label{subsec:scopedcaps} 

Interacting with files and sockets is a common task in daily programming. 
Many languages
provide scoped constructs, such as @with@ in Python or @try@-with-resources in Java, to
ensure that resources are properly released. In Scala, a typical pattern 
is to use a higher-order combinator @withFile@:
\begin{lstlisting}
def withFile[T](name: String)(op: File => T): T =  // library code
  val f = File.open(name)
  try op(f) finally f.close()  // grant access to f in op; revoke afterwards
withFile("a.txt"): f =>                            // user code
  f.write("Hello")             // while f is in scope, the file is open
  // f will be closed and go out of scope here
\end{lstlisting}

A key advantage of using scoped constructs is that the local variable @f@
guarantees the underlying file is open for the duration of the scope. 
Thus, @f@ embodies not only the file \emph{resource},
but also the \emph{capability}
\cite{DBLP:journals/cacm/DennisH66,DBLP:conf/asian/MillerS03} to operate on it.
Unfortunately, this guarantee holds only under conventional, first-order usage.
In impure higher-order languages such as Scala, programmers may inadvertently leak these capabilities in various ways:
\begin{lstlisting}
withFile("a.txt") { f => f }                         // $\color{red}\text{Leaked:}$ returned directly
withFile("a.txt") { f => throw f }                   // $\color{red}\text{Leaked:}$ thrown as exception
var f0: File; withFile("a.txt") { f => f0 = f }      // $\color{red}\text{Leaked:}$ stored in mutable vars
withFile("a.txt") { f => () => f.write("Hello") }    // $\color{red}\text{Leaked:}$ captured in closures
\end{lstlisting}

To prevent such leaks, several mechanisms have been proposed to ensure safe and ergonomic programming using capabilities in Scala. 
Broadly, they fall into two categories: 
(1) passing capabilities \emph{implicitly}, so that code never binds them to a named variable; 
and (2) \emph{explicitly} tracking
reachable/captured resources of function closures and data structures at the
type level.

\subsubsection{Remedy via Implicits.}
Using implicit function types \cite{DBLP:journals/pacmpl/OderskyBLBMS18}, we can write the same @withFile@ example 
without having the file handle variable in user code. 
For instance, the scoped file usage can be expressed alternatively:
\begin{lstlisting}
def withFile[T](name: String)(op: File ?=> T) = ... // ?=> for implicit function type
def write(text: String)(using f: File) = ...        // using for implicit arg list
withFile("a.txt") { write("Hello") }                // no explicit binding/passing of f
\end{lstlisting}
Here~@op: File ?=> T@ denotes an implicit function type where the @File@ argument is supplied implicitly by the compiler. 
Correspondingly, @write@ declares its @File@ parameter with a @using@ clause, and the file handle variable @f@ does not appear in the user code.

The implicit passing style is ergonomic, but it alone has notable limitations.
First, it does not scale to multiple files of the same type (like @File@), as the compiler cannot distinguish them, so in practice one still needs explicit names to avoid ambiguity.
Second, although the compiler supports implicit resolution, it does not enforce its use, 
meaning that programmers may still explicitly bind the file handle variable, even 
when using implicit function types.
\begin{lstlisting}
def mySummon[T](using c: T): T = c     // directly return the implicitly resolved argument
withFile("a.txt") { mySummon[File] }   // $\color{red}\text{Leaked!}$
\end{lstlisting}
Since implicits can be bypassed, the file handle @f@ can still escape the scope. 

As a result, the possibility of capability leakage remains. 
To obtain static safety guarantees against escaping, additional mechanisms are thus required.

\subsubsection{Prevent Leakage by Explicit Typing.}

Scala has recently introduced an experimental capture checker \cite{ScalaCC} to
prevent unintended escaping of resources. This checker implements a form of
\emph{descriptive alias tracking}, in which types are explicitly annotated with
variable names to represent the set of resources that may be captured or reached.
Such mechanisms have been formally studied under the names of 
capturing types (CT) \cite{DBLP:journals/toplas/BoruchGruszeckiOLLB23,DBLP:journals/pacmpl/XuBO24,DBLP:journals/pacmpl/XuBPO25}
and reachability types (RT) \cite{DBLP:journals/pacmpl/BaoWBJHR21,
DBLP:journals/pacmpl/WeiBJBR24}.
Although the systems discussed in the literature share foundational concepts,
they differ in important aspects such as the treatment of separation and the
handling of polymorphism. In this paper, we abstract over these differences by
focusing on a common core that suffices for our purposes. We illustrate this
shared foundation by demonstrating how it can safely encode the
scoped @withFile@ pattern.

\paragraph{Qualifiers: Sets of Variables.}

In both CT and RT, types are accompanied by qualifiers to describe which
variables may be captured or reached by a given term's result.
For example, when new files @fA@ and~@fB@ are opened,
their types are annotated with qualifiers containing their respective names:
\begin{lstlisting}
val fA = File.open("a.txt")       // fA: File$\trackset{fA}$
val fB = File.open("b.txt")       // fB: File$\trackset{fB}$
\end{lstlisting}

Qualifiers provide an over-approximation of the variables that a value may
capture or reach. For example, consider the variable @fC@, which may alias
either @fA@ or @fB@ depending on a runtime condition. The type of @fC@ can be
annotated with the qualifier @{fC}@, indicating that it is tracked by its own
name, or with @{fA,fB}@, reflecting the possibility that it aliases either @fA@
or @fB@.  This choice of qualifiers is possible by recording the potential
aliasing to both @fA@ and @fB@ in the typing context.
\begin{lstlisting}
val fC = if (...) fA else fB     // fC: File$\trackset{fC}$, or File$\trackset{fA,fB}$ $\dashv$ [$\ldots$, fC: File$\trackset{fA,fB}$]
\end{lstlisting}

\paragraph{Tracking in Higher-Order Functions.}

In higher-order languages such as Scala, functions may capture free variables or
manipulate first-class function closures. Both CT and RT provide mechanisms to
reason about such higher-order scenarios.
For example, consider an anonymous function that writes to a captured file 
handle @f@; the outermost qualifier of the function type should include @f@:
\begin{lstlisting}
val f = File.open("a.txt")   // f: File$\trackset{f}$
() => f.write("Hello")       // <anonymous>: (() => Unit)$\trackset{f}$
\end{lstlisting}

More interestingly, consider a function that accepts @f@ as a parameter and
returns the anonymous function capturing @f@; passing such a function to
@withFile@ would leak the file handle in a closure.
To enable the capture checker for detection, we annotate the
type of @f@ with~@^@, marking it as a tracked resource\footnote{%
With different semantics, the freshness marker~$\qfresh$ in RT-style
type signature achieves a similar role.} whose
aliasing the compiler will monitor.
The separation extension \cite{SeparationChecking}
further ensures that @f@
does not overlap with
captured free variables.
The result type @(() => Unit)@ is annotated
with @{f}@, reflecting the escape.
\begin{lstlisting}
def leakFile(f: File^) =    // leakFile: ((f: File^) -> (() ->^{f} Unit)) (CT style)
  () => f.write("Hello")    //           ((f: File$\trackfresh$) => (() => Unit)$\trackset{f}$)$\track\ $ (RT style)
\end{lstlisting}
When applying @leakFile@, the bound variable @f@ 
in types is substituted with the parameter qualifier:
\begin{lstlisting}
val fA = File.open("a.txt") // fA: File$\trackset{fA}$
val res = leakFile(fA)      // res: (() => Unit)$\trackset{fA}$ = [f$\:\mapsto\:$fA] (() => Unit)$\trackset{f}$
\end{lstlisting}

\paragraph{Polymorphism and Leakage Prevention.}

To specify the type of @withFile@, polymorphism over qualifiers is required.
In CT, the type variable @T@ is unqualified, and
any capturing must be tunneled using \emph{boxes}
\cite{DBLP:journals/pacmpl/BrachthauserSLB22}, which are automatically inferred
by the compiler. Under this scheme, @withFile@ prevents leakage because
the boxed @T@ cannot carry the tracked file handle out of scope:
\begin{lstlisting}
def withFile[T](name: String)(op: File^ => T): T
withFile("a.txt")(leakFile)  // $\color{red}\text{Error:}$ Ill-scoped unboxing
\end{lstlisting}
In RT \cite{DBLP:journals/pacmpl/WeiBJBR24}, the same guarantee is achieved
by explicitly quantifying over both a qualifier~@q@ and the type~@T@,
so that the result type~@T$\trackvar{q}$@ can only mention variables
visible to the caller, excluding the bound file handle:
\begin{lstlisting}
def withFile[T$\trackvar{q}$](name: String)(op: (File$\trackfresh$ => T$\trackvar{q}$)$\trackfresh$): T$\trackvar{q}$
withFile("a.txt")(leakFile)  // $\color{red}\text{Error:}$ Invalid subqual (RT)
\end{lstlisting}
For the purpose of this paper, we represent type annotations in the
RT style to align with the formal model of destructive effects \cite{DBLP:journals/corr/abs-2510-08939},
while our code remains compatible with the syntax of Scala capture checking.

\subsection{A Naive Step towards Typestate} \label{sec:naive}

While scope-based programming offers natural bounds on resource
lifetimes, it lacks the expressiveness required for certain scenarios. For
example, as illustrated by the lock example in \Cref{fig:firstlocks}, the
enforced LIFO discipline can inhibit desirable optimizations and flexible usage
patterns.

The same LIFO discipline also restricts flexibility when programming with files.
Thus, our objective is to support explicit @open@ and @close@ operations,
eliminating the need for scoped combinators like @withFile@.
At the same time, we need the type system to statically guarantee the safe
use of file operations with respect to file states.
To start with, we define two possible file states as classes:
\begin{lstlisting}
class OpenFile: ...                          // Not directly constructible
class ClosedFile: ...
\end{lstlisting}
With states defined, we can then define the operations that initialize files
and change file states:
\begin{lstlisting}
def newFile(name: String): ClosedFile = ...  // Construct a ClosedFile
def open(f: ClosedFile): OpenFile     = ...  // Transition a ClosedFile to OpenFile 
def close(f: OpenFile): ClosedFile    = ...  // Transition an OpenFile to ClosedFile
\end{lstlisting}
Last but not least, the operations that require files in open states:
\begin{lstlisting}
def read(f: OpenFile): String         = ...
def write(f: OpenFile, text: String): Unit = ...
\end{lstlisting}

With this API, we can express the same example previously demonstrated with
@withFile@, now using explicit typestate transitions. By distinguishing file
states with separate types, we statically guarantee that operations like
@write@ are only permitted for files in appropriate, opened states:
\begin{lstlisting}
val fNew = newFile("a.txt")
val fOpen = open(fNew)
write(fOpen, "Hello")                      // Good: Permitted only after opening the file
val fClosed = close(fOpen)
\end{lstlisting}

Nevertheless, this initial approach to typestate reasoning exhibits some significant shortcomings:

\paragraph{I: Lack of invalidation for outdated capabilities.}
After invoking @close(fOpen)@, the variable @fClosed@ represents the closed
state of the file. However, the original variable @fOpen@ remains valid in the
type system, allowing subsequent operations such as @write(fOpen, ...)@ to
type-check, even though the file has already been closed. This permits erroneous
use of stale references:
\begin{lstlisting}
val fClosed = close(fOpen)                 // fClosed supersedes fOpen
write(fOpen, "Hello")                      // $\color{red}\text{Stale}$ but type-correct
\end{lstlisting}

\vspace{-1ex}
\paragraph{II: Inability to verify resource identity.}
Because each state transition produces a new variable, the type system does not enforce that operations are performed on the intended resource. For example, it is possible to mistakenly operate on the wrong file without detection:
\begin{lstlisting}
val fA = newFile("a.txt"); val fB = newFile("b.txt")
val fOpen = open(fA)                       // a.txt is now open, but not b.txt
write(fOpen, "this should go to b.txt")    // $\color{red}\text{Unintended}$ but type-correct
\end{lstlisting}

\vspace{-1ex}
\paragraph{III: Poor ergonomics.}
This programming style requires explicit threading of stateful objects through sequences of function calls, resulting in cumbersome code.

\medskip
To overcome these limitations, we propose a flexible and expressive typestate
framework grounded in three key mechanisms. First, we introduce a destructive
effect system that statically tracks the revocation of capabilities. Second, we
employ path-dependent capabilities to preserve resource identity across state
transitions. Third, we leverage an A-normal form (ANF) transformation to enable
flow-sensitive implicit resolution. In the rest of this section, we
elaborate on these pillars.

\subsection{Pillar I: Flow-Sensitive Revocation of Capabilities}
\label{subsec:killeff}

To statically invalidate outdated capabilities, we introduce a flow-sensitive
\emph{destructive effect} system. This system uses the annotation
\lstinline|@kill($\ldots$)| on function result types to specify a set of
variables, free ones or arguments, whose use should be prohibited after the
function is applied:
\begin{lstlisting}
def newFile(name: String): ClosedFile^        = ...
def open(f: ClosedFile^): OpenFile^ @kill(f)  = ...
def close(f: OpenFile^): ClosedFile^ @kill(f) = ...
\end{lstlisting} 
The effect system sequentially tracks the accumulated set of killed variables,
represented as @$\JUSTK{\ldots}$@ below.
Invoking the effectful function @close@ amounts to extending it
with the parameter @fOpen@:
\begin{lstlisting}
val fClosed = close(fOpen)                 // $\JUSTK{\ldots, \texttt{fOpen}}$ extended
write(fOpen, "Hello")                      // $\color{red}\text{Error:}$ found using killed var fOpen
\end{lstlisting}

Unlike linear type systems \cite{DBLP:conf/ifip2/Wadler90,DBLP:journals/pacmpl/SpiwackKBWE22}, where any
function using a linear resource must consume it, the \lstinline|@kill| 
annotation in our system provides selective, opt-in revocation.
Functions that merely use capabilities without revoking them (\eg, @write@)
do not induce any destructive effect.
This design enables seamless integration with
imperative and higher-order constructs:
\begin{lstlisting}
val messages: Array[String] = ...
val fOpen = open(newFile("a.txt"))         // open is effectful
for (msg <- messages) write(fOpen, msg)    // loop of writes is free of kill!
close(fOpen)                               // close is effectful
\end{lstlisting}

\paragraph{Foundation: Reachability Types and Transitive Disjointness.}

When capabilities are not required to be linear, sharing and
aliasing become possible. In this setting, our effect system must ensure that
all potential aliases of a killed capability are also invalidated to provide
strong static guarantees.
Our model reasons about aliases with reachability types \cite{DBLP:journals/corr/abs-2510-08939}.
For example, consider the following scenario, where @fC@ may alias
either @fA@ or @fB@, as recorded in the typing context:
\begin{lstlisting}
val fA = open(newFile("a.txt")); val fB = open(newFile("b.txt"))
val fC = if (...) fA else fB               // context: [$\ldots$, fC: OpenFile$\trackset{fA,fB}$]
close(fA)                                  // $\JUSTK{\ldots,\texttt{fA}}$ extended
write(fC, "maybe to a.txt")                // $\color{red}\text{Error:}$ found using killed var fA
\end{lstlisting}

After closing @fA@, subsequent writes to @fC@ are unsafe, as @fC@ may alias the
now-closed file.  On the other hand, the accumulated set of killed variables is
extended by only @fA@, not @fC@. To prevent this misuse, we require that the
qualifier of any used term be \emph{transitively disjoint} from the killed set.
In this example, the typing context reveals that @fC@ may reach both @fA@ and
@fB@; thus, the separation check fails due to the presence of @fA@ in the
transitive closure of @fC@'s qualifier:
\vspace{-.5ex}
\[
\footnotesize
\qsat{\texttt{\color{dark-cyan}fC}} \cap \qsat{{\color{eff}\KILL}}
\quad=\quad\texttt{\color{dark-cyan}\{fA,fB,fC\}} \cap \texttt{\color{eff}\{\ldots,fA\}}
\quad=\quad\texttt{\{fA\}}
\quad{\;\color{red}\nsubseteq\;}\quad \qbot
\vspace{-.5em}
\]
To sum up, closing @fA@ disables access to both @fA@ and any variable that
may reach it, such as @fC@, while leaving @fB@ unaffected. In contrast, closing
@fC@ disables access to all three variables: @fA@, @fB@, and @fC@.

An alternative perspective on the effect separation check is provided by
continuation-passing style (CPS) \cite{DBLP:journals/mscs/DanvyF92}. In this formulation, APIs such as
@close@ and @write@ are extended to accept an explicit continuation parameter.
Without any effect system, reachability types alone can prevent reusing
revoked @OpenFile@ handles by enforcing transitive disjointness between
the continuation @k@ and the handle~@f@
when both are annotated with the fresh qualifier $\qfresh$:
\begin{lstlisting}
// closeCPS: (f: OpenFile$\trackfresh$) => (k: (ClosedFile$\trackfresh$ => NoReturn)$\trackfresh$) => NoReturn
closeCPS(fA){ fA_ => writeCPS(fC, "maybe to a.txt"){...} }   // $\color{red}\text{Error:}$ f,k overlap on {fA}
\end{lstlisting}

Back in direct-style programming, where there is no explicit notion of
continuations, we employ effect tracking and effect separation instead.
In CPS, borrowing a resource implicitly becomes permanent because the continuation never returns,
which makes a borrow behave effectively like a move.
Our direct-style kill effects mirror this observation by ensuring that once a capability is revoked, all reachable aliases are invalidated.

\paragraph{Notation.}

We represent the types of functions that kill their
arguments using the arrow @=!>@ and its implicit variant @?=!>@.
The signatures of @open@ and @close@ can be simplified accordingly.
\begin{lstlisting}
type =!>[S, T]  = (c: S) =>  T @kill(c)      // arrow type     that kills arg c
type ?=!>[S, T] = (c: S) ?=> T @kill(c)      // implicit arrow that kills arg c
// open: ClosedFile$\trackfresh$ =!> OpenFile$\trackfresh$; close: OpenFile$\trackfresh$ =!> ClosedFile$\trackfresh$
\end{lstlisting}

\subsection{Pillar II: Relating Capabilities and Objects by Path-Dependent Types}\label{subsec:pathdep}

While the effect system ensures that new states of an object supersede previous
ones, it does not distinguish between states of different objects. To
illustrate this limitation, consider a variant of the @withFile@ combinator,
named @ensureClosed@, which operates over the state class @ClosedFile@:
\begin{lstlisting}
def ensureClosed(name: String)(op: ClosedFile^ =!> ClosedFile^): Unit =
  op(newFile(name)); ()
\end{lstlisting}
Here, the type of @op@ enforces that a fresh @ClosedFile@ is returned, thereby
requiring that any file opened within the scope of @ensureClosed@ must be closed
before the function returns:
\begin{lstlisting}
ensureClosed("a.txt"): f =>
  val fOpen = open(f)
  write(fOpen, "Hello")
  close(fOpen)           // Error if omitted
\end{lstlisting}

However, the type system does not guarantee that the @ClosedFile@ returned by
@op@ is the same file that was originally provided. For example, the type
checker would accept an instance of @op@ that simply returns a newly
created, unopened file, rather than the intended one:
\begin{lstlisting}
ensureClosed("a.txt"): f => 
  val fOpen = open(f)
  newFile("b.txt")       // $\color{red}\text{Unintended}$ but type-correct
\end{lstlisting}
More fundamentally, even if @newFile@ is removed from the API, a
programmer can still circumvent the intended guarantees by reusing a
@ClosedFile@ obtained from an outer @ensureClosed@ block.
A rigorous mechanism relating capabilities and object identities is necessary.

\paragraph{Why not Reachability/Capturing Types.}

Although descriptive alias tracking mechanisms offer promising ways to reason
about resources, they do not address this identity problem.
As illustrated by the signatures of @open@ and @close@ above, these APIs
always revoke the provided capability and generate a fresh one; at the type
level, no relationship is maintained between the input and output capabilities.
Attempting to relate them would result in the returned capability being
immediately invalidated by the kill effect.
Furthermore, both RT and CT are inherently over-approximations: a qualifier
@{f}@ indicates that a capability may, but does not necessarily, refer to @f@.
Consequently, these systems cannot provide the desired safety properties (\Cref{sec:safety-properties}) here.

\paragraph{Our Solution: Path-Dependent Capabilities.} 

Independent of reachability types and effects that govern the lifetime of
capabilities, we leverage \emph{path-dependent types} from Dependent Object
Types (DOT) \cite{DBLP:conf/oopsla/RompfA16,DBLP:conf/birthday/AminGORS16}
to track capability identities.
In Scala, a class may declare \emph{abstract type members}: types that are
left unspecified in the class definition and whose concrete identity is
determined per object instance.
Because each instance carries its own copy of these types,
they are effectively existential: 
for two variables @a@ and @b@ of the same class, 
the \emph{path-dependent} types @a.T@ and @b.T@ are considered distinct by the
compiler.
We exploit this property to represent files as a
unified class @File@, with the two possible states as abstract type members:
\begin{lstlisting}
class File:
  type IsClosed          // abstract type members
  type IsOpen            // for f of type File, there are caps: f.IsClosed and f.IsOpen
\end{lstlisting}

Crucially, for any two distinct variables @f@ and @g@ of type @File@, the
corresponding path-dependent types @f.IsClosed@ and
@g.IsClosed@ are also distinct and cannot be confused by the type system. This
property enables us to define file APIs in a path-dependent manner:
\begin{lstlisting}
def openDep(f: File, c: f.IsClosed^): f.IsOpen^ @kill(c)   = ...
def closeDep(f: File, c: f.IsOpen^):  f.IsClosed^ @kill(c) = ...
def readDep(f: File, c: f.IsOpen^):   String               = ...
def writeDep(f: File, s: String, c: f.IsOpen^): Unit       = ...
\end{lstlisting}
In these APIs, @f@ denotes the file \emph{resource}, while @c@ is a
\emph{path-dependent capability} whose type is prefixed by the specific variable
@f@. The transition functions @openDep@ and @closeDep@ consume (kill) their
input capabilities and return fresh ones, all associated with the same file via
the path prefix @f@.

The scoped combinator @ensureClosedDep@ additionally provides the initial capability and
enforces that the returned capability corresponds to the same file, thereby
guaranteeing that resources are properly closed upon exiting the scope and
preventing confusion between object identities:
\begin{lstlisting}
def ensureClosedDep(name: String)(op: (f: File) => f.IsClosed^ =!> f.IsClosed^): Unit =
  ...
ensureClosedDep("a.txt"): f => cInit =>
  val cOpen = openDep(f, cInit)
  writeDep(f, "Hello", cOpen)
  closeDep(f, cOpen)             // Error if omitted
ensureClosedDep("a.txt"): f1 => cInit1 =>     
  val cOpen1 = openDep(f1, cInit1)
  ensureClosedDep("b.txt"): f2 => cInit2 =>
    closeDep(f1, cOpen1)         // $\color{red}\texttt{Error:}$ expect f2.IsClosed, got f1.IsClosed
\end{lstlisting}
Crucially, safe use of such APIs requires disciplined encapsulation. %
To keep the type members of @File@ abstract, 
it is critical that we only introduce files via @ensureClosedDep@,
whereas common constructors could instantiate capabilities with arbitrary types. 
This can be enforced with appropriate encapsulation techniques (e.g. @private@
constructors) orthogonal to our presentation.

\paragraph{Bundling Resources and Capabilities as $\Sigma$.}

A final challenge remains in adapting the @newFile@ operation to the
path-dependent capability style. Unlike @ensureClosedDep@, which supplies both
the file and its initial @IsClosed@ capability as separate, yet dependent,
arguments within a scope, the imperative @newFile@ must return both the file
object and its associated capability together, while preserving their type-level
dependency. This necessitates a mechanism for simultaneously constructing and
returning a resource and its path-dependent capability in a type-safe manner.

A natural solution to this problem is dependent pairs, also known as $\Sigma$ types.
To achieve this, we define a @trait@, Scala's analogue of an interface, 
named @Sigma@:

\noindent
\lstinline|trait Sigma {|
$\underbrace{\text{\lstinline|type A; type B|}}_{\text{abstract types}}$\lstinline|;|
$\underbrace{\text{\lstinline|val a: A; val b: B|}}_{\text{fields typed by A, B}}$
\lstinline| }|

To use it as the result type of @newFileSigma@, @Sigma@ can be refined
with concrete, dependent @A@ and @B@.
We instantiate @A@ with the type of resources, @File@ here.
Crucially, we instantiate @B@ as the type of the path-dependent capabilities by
referring to the value field @a@ within @Sigma@:
\begin{lstlisting}
def newFileSigma(name: String): Sigma { type A = File; type B = a.IsClosed^ } = ...
\end{lstlisting}

Crucially, @Sigma@ should be understood as a \emph{transient} wrapper for bundling
resources and capabilities, backed by specialized compiler support,
but not a dependent type data structure with reachability tracking, which
is beyond the scope of this work.
Once returned from @newFileSigma@, the result @sigma@ needs immediate unpacking
to maintain sound reachability tracking.
To preserve the type-level dependency, we need to ascribe the field, @a@, using
singleton types \cite{DBLP:conf/oopsla/OderskyZ05}:
\begin{lstlisting}
val sigma = newFileSigma("a.txt") 
val f: sigma.a.type = sigma.a    // sigma.a.type: singleton type of sigma.a
val c = sigma.b                  // c: f.IsClosed$\trackset{c}$
\end{lstlisting}

\subsection{Pillar III: Flow-Sensitive Introduction of Capabilities}\label{subsec:implicitintro}

While the combination of destructive effects and path-dependent capabilities
yields strong safety guarantees, programming directly with these mechanisms 
can be verbose and unwieldy. In this section, we present a series of techniques 
to improve the ergonomics of typestate programming, enabling 
more concise and user-friendly code without compromising safety.

\paragraph{Implicit Resolution.}
To enable implicit argument resolution \cite{DBLP:journals/pacmpl/OderskyBLBMS18} for our APIs,
we can declare the capability argument in a separate argument list led by
the @using@ keyword:
\begin{lstlisting}
def openImp(f: File)(using c: f.IsClosed^): f.IsOpen^ @kill(c) = ...
\end{lstlisting}
Or, more concisely, using the notations for implicit arrows and destructive arrows:
\begin{lstlisting}
def openImp(f: File):  f.IsClosed^ ?=!> f.IsOpen^    = ...  // ?=!>: implicit + kill
def closeImp(f: File): f.IsOpen^   ?=!> f.IsClosed^  = ...
def readImp(f: File):  f.IsOpen^   ?=>  String       = ...  // ?=>: implicit only
def writeImp(f: File, s: String): f.IsOpen^ ?=> Unit = ...
\end{lstlisting}

With these APIs leveraging implicit resolution, capabilities no longer require
explicit passing. However, implicit instances
must still be declared explicitly, which introduces additional complexity. In
particular, unpacking the bundled @Sigma@ type requires singleton type
ascription, and careful scoping is needed to disambiguate multiple live capabilities 
of the same type, including revoked ones.
Without additional mechanisms, programmers must manually structure scopes to maintain
unambiguous implicit resolution, which is undesirable.

\paragraph{$\Sigma$-Guided ANF Transformation.}

To facilitate the ergonomic, flow-sensitive introduction of path-dependent 
capabilities encapsulated within @Sigma@ types, we employ a type-directed A-normal form (ANF)
transformation \cite{DBLP:conf/icfp/RompfMO09}. Specifically, for any non-tail
expression of type @Sigma@, the transformation restructures the continuing
computation into a new block. Within this block, the first field @a@ is
extracted and ascribed a singleton type, while the second field @b@ is
declared as an implicit candidate. This block-based approach ensures that the
newly introduced implicit has the highest precedence in subsequent
resolution, thereby eliminating ambiguity and supporting reliable capability
inference:

\noindent
\begin{tabular}{m{.36\linewidth}lm{.5\linewidth}}
\begin{lstlisting}
val f = newFileSigma()
openImp(f)
\end{lstlisting}
& $\Longrightarrow\quad$ &
\begin{lstlisting}
val sigma_0 = newFileSigma() 
{
  implicit val sigma_0_imp = sigma_0.b 
  val f: sigma_0.a.type = sigma_0.a
  openImp(f)   // inferred using sigma_0_imp
}
\end{lstlisting}
\end{tabular}

More generally, other APIs can be refactored to return a @Sigma@ type, enabling
their use to benefit from the ANF transformation described above. For example, a
variant of @open@ may return the new capability as the second field of a
@Sigma@, with the first field instantiated as @Unit@:
\begin{lstlisting}
def openSigma(f: File): f.IsClosed^ ?=!> Sigma { type A = Unit; type B = f.IsOpen^ } = ...
\end{lstlisting}

\paragraph{Implicit $\Sigma$-Lifting.}
To further streamline the construction of @Sigma@ results, we introduce implicit
$\Sigma$-lifting. This mechanism is particularly beneficial for 
combinators such as @ensureClosedSigma@, which require the callback @op@ to
return both a data value of type @T@ and an @IsClosed@ capability witness:
\begin{lstlisting}
def ensureClosedSigma[T](name: String)
  (op: (f: File) => f.IsClosed^ ?=!> Sigma { type A = T; type B = f.IsClosed^ }): T = ...
\end{lstlisting}
When returning the result read from the file, it is natural for users to simply
return the string variable @text@. However, the expected return type is a
dependent pair (@Sigma@), requiring the result and a capability to be
bundled together. To reconcile this mismatch, the compiler automatically lifts
the return value into the first field @a@ of the @Sigma@ pair, while the second
field @b@ is populated by implicitly summoning the appropriate capability. This
implicit $\Sigma$-lifting mechanism ensures that the returned value conforms to
the required dependent pair type without additional user intervention:\footnote{
Due to current limitations in Scala regarding curried dependent implicit
function types, the implicit parameter \lstinline|c| must be explicitly bound
and passed.  This restriction is incidental to our approach; for clarity, we
omit it in subsequent examples.  }

\noindent
\begin{tabular}{m{.42\linewidth}@{}c@{}m{.52\linewidth}@{}}
\begin{lstlisting}
ensureClosedSigma("a.txt"): f => c ?=>
  openSigma(f)(using c)
  val text = readSigma(f)
  closeSigma(f)
  text  // needs lifting
\end{lstlisting}
& $\Longrightarrow\ $ &
\begin{lstlisting}
ensureClosedSigma("a.txt"): f => c ?=>
  ...
  new Sigma:
    type A = String
    type B = f.IsClosed^ 
    val a = text
    val b = summon[f.IsClosed]  //<-closeSigma
\end{lstlisting}
\end{tabular}

Given the ANF transformation creating a new scope for @closeSigma@,
the summon can locate the most recent, live capability for @f.IsClosed@.

\paragraph{Notation.}

As a dual to implicit function types (@?=>@), which receive implicit arguments,
the $\Sigma$-guided ANF transformation enables the implicit return of results,
thus complementing the flow-sensitive revocation of capabilities in \Cref{subsec:killeff}.
To make this duality explicit, we introduce the arrow notation @?<=@ as an
alternative to @Sigma@:
\begin{lstlisting}
type ?<=[B1, A1] = Sigma { type A = A1; type B = B1 } 
// openSigma: (f: File) => f.IsClosed$\trackfresh$ ?=!> f.IsClosed$\trackfresh$ ?<= Unit
\end{lstlisting}

Going further, for functions such as @openSigma@ that perform only state
transitions and do not produce an explicit output, we introduce the combined
arrow notation @?=!>?@. This notation succinctly expresses three key aspects:
(1) the function receives an implicit capability argument, (2) the capability is
revoked via a destructive effect, and (3) a new implicit capability is returned.
This abstraction streamlines the specification of typestate transitions,
improving both the clarity and conciseness of API signatures. %
\begin{lstlisting}
type ?=!>?[S1, S2] = (c: S1^) ?=!> ((S2^) ?<= Unit)
// openSigma:  (f: File) => f.IsClosed ?=!>? f.IsOpen
// closeSigma: (f: File) => f.IsOpen   ?=!>? f.IsClosed
\end{lstlisting}

\subsection{Summary}

In this section, we have developed a generic typestate framework by
progressively extending scoped capability-based file programming with three
key mechanisms:
\begin{itemize}[leftmargin=*]
\item \textbf{Flow-sensitive revocation} (\Cref{subsec:killeff}):
  a destructive effect system that statically invalidates outdated capabilities,
  ensuring that stale references cannot be used after a state transition.
\item \textbf{Path-dependent capabilities} (\Cref{subsec:pathdep}):
  abstract type members that relate each capability to a specific object instance,
  preventing confusion between the states of different resources.
\item \textbf{$\Sigma$-guided ANF transformation} (\Cref{subsec:implicitintro}):
  a type-directed rewriting that automatically introduces and resolves
  capabilities via implicits, eliminating manual threading.
\end{itemize}
To illustrate how these pieces fit together, the file API developed
throughout this section is summarized below, alongside a usage example
in its final, ergonomic form:
\begin{lstlisting}
// API
def newFile(name: String): Sigma { type A = File; type B = a.IsClosed^ }
def open(f: File):             f.IsClosed ?=!>? f.IsOpen    // receive, revoke, return
def close(f: File):            f.IsOpen   ?=!>? f.IsClosed
def read(f: File):             f.IsOpen   ?=>   String
def write(f: File, s: String): f.IsOpen   ?=>   Unit
// Usage
val f = newFile("a.txt")       // f: File, f.IsClosed introduced
open(f)                        // f.IsClosed consumed, f.IsOpen introduced
write(f, "Hello")              // f.IsOpen summoned
close(f)                       // f.IsOpen consumed, f.IsClosed introduced
\end{lstlisting}
In the rest of the paper, we present additional case studies
to illustrate the use of our framework across
diverse domains and elaborate on the underlying design and implementation. %
\section{Case Studies} \label{sec:casestudy}
In this section, we present additional case studies to demonstrate
our programming model.
To keep the presentation focused on key concepts, 
we defer low-level implementation details to the appendix (\Cref{appendix:case-study}).
All definitions and examples presented in this section are fully supported and 
type checked in our prototype implementation \cite{artifact}.

\subsection{Table Locking} \label{sec:lockagain}

\begin{figure}[b]
\begin{lstlisting}[style=impl,name=tablelock]
trait Lock: |\label{line:locktrait}|
  type IsHeld      // lock held,     resource usable 
  type IsReleased  // lock released, resource unusable

class Table private[...](n: Int) extends Lock:  // constructor private to API package |\label{line:tableclass}| 
  private val data = ...  // an indexable data storage
  // ... table lock fields ...
  class Row private[...](m: Int) extends Lock:  // constructor private to API package |\label{line:rowclass}| 
    private val row = data(m)  // mth row of table
    // ... row lock fields ...   

object Table: 
  def apply(n: Int): Sigma { type A = Table; type B = a.IsReleased^ } = ...  // factory method |\label{line:tableapply}|

extension (table: Table) |\label{line:tableextension}|
  def lock():   table.IsReleased ?=!>? table.IsHeld = ...  // acquire the lock of table |\label{line:tablelock}|
  def unlock(): table.IsHeld ?=!>? table.IsReleased = ...  // release the lock |\label{line:unlock}|
  def locateRow(n: Int): table.IsHeld^ ?=> Sigma { type A = table.Row; type B = a.IsReleased^ } = ...
    // locate nth row of table |\label{line:locaterow}| 
  def lockRow(row: table.Row): table.isHeld^ ?=> row.IsReleased ?=!>? row.IsHeld = ... |\label{line:rowlock}|
    // acquire the lock of row 

extension (row: Table#Row)
  def unlock(): row.IsHeld ?=!>? row.IsReleased = ...        // release row lock |\label{line:rowunlock}|
def computeOnRow(row: Table#Row): row.IsHeld^ ?=> ... = ...  // compute on row |\label{line:computeonrow}|
\end{lstlisting}
\caption{Table Locking Definitions and API}
\label{fig:lockdefsandapi}
\end{figure}

Our first case study revisits the imperative table locking example 
from \Cref{fig:flowlock1} in \Cref{sec:intro}.
\Cref{fig:lockdefsandapi} presents one possible implementation, including the @Table@ definitions and the corresponding 
API. To track the lock status of tables and rows, we define a mixin @Lock@ (line \ref{line:locktrait})
containing a type member for each state. 
Both the @Table@ (line \ref{line:tableclass}) and its nested class 
@Row@ (line \ref{line:rowclass}) extend @Lock@, so that each table and row carry their own lock states. 
Because @Row@ is defined inside
@Table@, every @Row@ instance is statically tied to its parenting @Table@.

The factory method on line \ref{line:tableapply} creates a new @Table@ together with its initial capability @isReleased@, packaged as a @Sigma@ pair.
The @Table@ itself
is returned explicitly, while the @isReleased@ capability is returned implicitly. 
Line \ref{line:tableextension} defines 
a collective extension for @Table@ instances, allowing additional methods to
be invoked in the standard object-oriented style, such as @table.lock()@.

Methods that perform typestate transitions use the composite arrow @?=!>?@. 
For example, @lock@ (line \ref{line:tablelock}) acquires the lock on a @Table@ by consuming the 
@table.IsReleased@ capability and returning a @table.IsHeld@ capability.
Then, operations that need @Table@ to be in a specific state  
are implicitly parameterized by the corresponding path-dependent capability. 
Retrieving a row (line \ref{line:locaterow}) requires the @Table@ lock to be held, 
so it requires the @table.IsHeld@ capability, and returns 
a @Row@ value which depends on the same @Table@ instance, together with its @isReleased@ capability.

Notably, %
to lock a @Row@, the enclosing @Table@ must already be locked, but unlocking 
a @Row@ does not require so.
Accordingly, @lockRow@ (line \ref{line:rowlock}) operates on a specific
@table@ instance and requires the @table.isHeld^@ capability, whereas @unlock@
(line \ref{line:rowunlock}) requires no specific table. This independence is
expressed using the \emph{type projection} @Table#Row@
\cite{DBLP:conf/oopsla/OderskyZ05}, which is used to refer to arbitrary @Row@
objects.  

\subsection{DOM Trees} \label{sec:domtree}
The examples so far have been modeled as finite-state machines,
while our programming model is also capable of 
tracking typestate defined by a context-free grammar. 
To demonstrate this expressiveness, we sketch a stateful API for constructing DOM trees, 
where the typestate corresponds to a list of currently open brackets:\footnote{
Due to Scala limitations with curried implicit dependent function types,
the first line should bind the capability explicitly,
\ie, \lstinline|dom => c =>|. As before, we omit \lstinline|c| as it is incidental to our approach.}

\begin{lstlisting}[escapechar=|]
makeDOM: dom => |\label{line:elidedparamDOM}|
  dom.open(DIV())
  // ... adding text to DIV ... 
  dom.close(P())    // $\color{red}\text{Error:}$ state is [DIV] not [P, ...]
  dom.close(DIV()) 
  dom.close(DIV())  // $\color{red}\text{Error:}$ state is [], not [DIV, ...]
\end{lstlisting}

\Cref{fig:domdefsandapi} depicts a minimal set of interfaces 
required to track such typestate. 
Tracking a list of open brackets necessitates expressing lists of DOM elements
as types.
We first introduce a sum type @Elem@ (line \ref{line:elemtrait}), with one variant per DOM node. 
Subsequently, we define a \emph{heterogeneous type list} \cite{DBLP:conf/haskell/KiselyovLS04} @EList@  (line \ref{line:elist})
of @Elem@ types. The @DOM@ class (line \ref{line:domclass}) possesses
a higher-kinded type member @Elems@ parameterized by an @EList@.
Different @DOM@ states are hence different @EList@ parameters to @Elems@.

\begin{figure}[h]
\begin{lstlisting}[style=impl] 
trait Elem; class DIV() extends Elem; class P() extends Elem; ... // other elements |\label{line:elemtrait}|
trait EList; class ENil extends EList |\label{line:elist}|
class ::[E <: Elem, L <: EList] extends EList                     // :: can be used as infix 

class DOM private[...]() { type Elems[L <: EList];  ... /* other fields */ }  |\label{line:domclass}|

extension (dom: DOM) 
  def open[E <: Elem, L <: EList](elem: E):  dom.Elems[L] ?=!>? dom.Elems[E :: L] = ...  |\label{line:domopen}|
  def close[E <: Elem, L <: EList](elem: E): dom.Elems[E :: L] ?=!>? dom.Elems[L] = ... |\label{line:domclose}|
  def text[E <: Elem, L <: EList](elem: E, s: String): dom.Elems[E :: L]^ ?=> Unit = ... |\label{line:domtext}|

def makeDOM(body: (dom: DOM) => (dom.Elems[ENil]^) =!> (dom.Elems[ENil]^) ?<= Unit): Unit = ... |\label{line:makeDOM}|
\end{lstlisting}
\caption{DOM Definitions and API}
\label{fig:domdefsandapi}
\end{figure}

The @DOM@ object tracks two kinds of state changes: opening and closing elements. 
Opening an element @E@ transitions the @EList@ state from @L@ to @E :: L@, prepending @E@ to the list. 
Conversely, closing an element (line \ref{line:domclose}) performs the dual operation, dropping @E@.
This mechanism statically enforces correct bracketing, that is, each opened element must be closed in the reverse order. 

\begin{figure}[t] 
\begin{subfigure}[t]{0.45\linewidth}
\begin{lstlisting}[style=impl,numbers=none]
makeDOM: dom => 
  dom.open(DIV()) 
  dom.close(DIV())         
  dom.close(DIV())  // $\color{red}\texttt{Error}$
\end{lstlisting}
\vspace{-1ex}
\caption{$\color{red}\text{Error}$ on line 4, due 
to use of killed variable with type \lstinline@dom.Elems[DIV :: ENil]@}
\label{fig:domerr1}
\end{subfigure}\hspace{0.7cm}
\begin{subfigure}[t]{0.45\linewidth}
\begin{lstlisting}[style=impl,numbers=none]
makeDOM: dom => 
  dom.open(DIV())
  dom.close(DIV())
  dom.close(HEAD())  // $\color{red}\texttt{Error}$
\end{lstlisting}
\vspace{-1ex}
\caption{$\color{red}\text{Error}$ on line 4, since 
no implicit found of type \lstinline@dom.Elems[HEAD :: ...]@}
\label{fig:domerr2}
\end{subfigure}
\caption{Errors caught by the DOM API}
\vspace{-1ex}
\end{figure}

To introduce @DOM@ objects, the API provides a higher-order function @makeDOM@ 
(line \ref{line:makeDOM}), analogous to @withFile@ to ensure that the @DOM@ tree is 
fully bracketed. Its @body@ parameter takes a @DOM@ object @dom@ and a @dom.Elems[ENil]@ capability, and must return the same capability using @Sigma@, 
guaranteeing that all opened elements are eventually closed. %
This design detects several kinds of errors at compile-time, such as double element closure (\Cref{fig:domerr1})
or closing an element (\Cref{fig:domerr2}) that is never opened.

\paragraph{DOM API Example.}
\begin{figure}[t]
\begin{lstlisting}[style=impl] 
// creates </tr><tr>
def nextTR[L <: EList](t: DOM): (t.Elems[TR :: L] ?=!>? t.Elems[TR :: L]) = |\label{line:nextTR}|
  t.close(TR()); t.open(TR())

// creates <td>p._1</td> <td>p._2</td>
def twoCells[L <: EList](t: DOM, f: String, s: String): t.Elems[TR :: L] ?=!>? t.Elems[TR :: L] = |\label{line:twoCells}|
  t.open(TD()); t.text(TD(), f); t.close(TD());
  t.open(TD()); t.text(TD(), s); t.close(TD())
 
def main() = makeDOM { dom => 
  dom.open(TABLE()); dom.open(TBODY()); dom.open(TR()) 
  val response = await(fetch("/api/logs").toFuture) // fetch log file
  val reader = response.body.getReader()            // get log file reader
  type TStart = TR :: TBODY :: TABLE :: ENil 

  def readAll(dom: DOM): dom.Elems[TStart] ?=!>? dom.Elems[ENil] = |\label{line:readAll}|
    val chunk = await(reader.read().toFuture)       // read chunk from log file 
    if (!chunk.isDone) then  
      val line = chunk.value                        // get line from chunk 
      twoCells(dom, line.timeStamp, line.message); nextTR(dom) 
      readAll(dom) 
    else 
      dom.close(TR()); dom.close(TBODY()); dom.close(TABLE()) // close table
  readAll(dom) } 
\end{lstlisting}
\caption{Building an HTML table containing log file lines using the DOM API.}
\label{fig:domexample}
\end{figure}

\Cref{fig:domexample} demonstrates how this API enables precise control over
opening and closing @DOM@ trees.  The example builds an HTML table where each
row is a separate line of a log file, assuming a standard HTTP
request API capable of fetching data asynchronously.

Before @main@, auxiliary methods @nextTR@ (line \ref{line:nextTR}) and
@twoCells@ (line \ref{line:twoCells}) are defined.  Using manual open and close
operations, @nextTR@ creates a non-bracketed row transition @</tr><tr>@, and
@twoCells@ constructs bracketed columns.  The two functions both use @?=!>?@
to indicate that they perform state transitions on the argument.

The @main@ function first opens the HTML table elements before fetching the log
file.  It then defines the recursive method @readAll@ (line \ref{line:readAll}),
which repeatedly reads from the log file and creates a new table row for each
line of the log file.  The @readAll@ method takes a @DOM@ object and transitions its
typestate from @TStart@, the initial state of the table, to @ENil@, representing
a fully closed tree.  Each iteration reads a new chunk via the asynchronous
reader.  If the chunk is not yet complete (@isDone@), it extracts the log line
and uses @twoCells@ to emit two table cells (timestamp and message), and then calls
@nextTR@ to start a new row.  The recursive call then proceeds with the updated
typestate produced by @nextTR@, which type-checks.  In the else branch,
@readAll@ closes the table and produces the final typestate @dom.Elems[ENil]@.
At the end of the @makeDOM@ block, @readAll@ is invoked, ensuring that
all elements are properly closed and the resulting DOM tree is well-bracketed.

\subsection{Session Types} \label{sec:sessiontypes}
Session types \cite{DBLP:conf/concur/Honda93, DBLP:conf/parle/TakeuchiHK94, DBLP:conf/esop/HondaVK98} 
are a type-based discipline for specifying and verifying communication protocols, ensuring type safety and adherence to protocols. 
In particular, \emph{binary session types} regulate communication between two parties, each associated with a type describing its protocol:
\begin{lstlisting}
// protocol of this channel endpoint: Send[String, Send[String, Recv[Int, End]]]
chan.send("Hello"); chan.send("World"); println(chan.recv() + 20)
\end{lstlisting}

To ensure communication safety, session-typed channel endpoints adhere to the 
property of \emph{duality} that every
message sent by one endpoint is received by the other. The dual of the protocol in the above 
example is @Recv[String, Recv[String, Send[Int, End]]]@, which the other endpoint of the channel
must be typed upon.

\begin{figure}[t] 
\begin{subfigure}[t]{0.46\linewidth}
\begin{lstlisting}[style=impl, name=sessiontypes]
trait Local  // Extended by types below |\label{line:sessiontrait}|
class Send[B, T <: Local] ... |\label{line:sendclass}|
class Recv[B, T <: Local] ... |\label{line:recvclass}|
class Branch[L <: Local, R <: Local] ... |\label{line:branchclass}|
class Select[L <: Local, R <: Local] ... |\label{line:selectclass}|
class End ... |\label{line:endclass}|
class Rec[T <: Local] ... |\label{line:recclass}|
class Var[N <: Int] ... |\label{line:varclass}|

trait PList |\label{line:sessionplist}|
class PNil extends PList
class ::[T <: Local, L <: PList] extends PList
\end{lstlisting}
\end{subfigure}
\hspace{0.7cm}
\begin{subfigure}[t]{0.46\linewidth}
\begin{lstlisting}[style=impl,xleftmargin=1em, name=sessiontypes]
class Chan private[...](): // Channel |\label{line:chanclass}|
  type PCap[E <: PList, T <: Local] |\label{line:chanpcap}|
  // ... more fields

type Dual[T <: Local] <: Local = T match |\label{line:dualtype}|
  case Send[b, t] => Recv[b, Dual[t]]
  case Recv[b, t] => Send[b, Dual[t]]
  case Branch[l, r] => Select[Dual[l], Dual[r]]
  case Select[l, r] => Branch[Dual[l], Dual[r]]
  case End => End
  case Rec[t] => Rec[Dual[t]]
  case Var[n] => Var[n]
\end{lstlisting}
\end{subfigure}
\begin{subfigure}{\linewidth}
\smallskip
\begin{lstlisting}[style=impl,xleftmargin=5ex,name=sessiontypes]
object Chan:
  def apply[T <: Local](): // factory method, creates two dual channels |\label{line:chanapply}|
    ( Sigma { type A = Chan; type B = a.PCap[PNil, T]^ },
      Sigma { type A = Chan; type B = a.PCap[PNil, Dual[T]]^ }) = ...

// type parameter bounds omitted: E <: PList, and T, L, R all <: Local 
extension (chan: Chan)
  // basic Channel Operations
  def send[B, E, T](x: B): chan.PCap[E, Send[B, T]] ?=!>? chan.PCap[E, T] = ... // sends x: B |\label{line:chansend}|
  def recv[B, E, T](): (chan.PCap[E, Recv[B, T]]^) ?=!> ((chan.PCap[E, T]^) ?<= B) = ... // recv B |\label{line:chanrecv}|
  def close[E](): chan.PCap[E, End] ?=!> Unit = ... |\label{line:chanclose}|
  // channel choice
  def left[E, L, R]():  chan.PCap[E, Select[L, R]] ?=!>? chan.PCap[E, L] = ... // choose left |\label{line:chanleft}|
  def right[E, L, R](): chan.PCap[E, Select[L, R]] ?=!>? chan.PCap[E, R] = ... // choose right |\label{line:chanright}|
  def branch[E, L, R, F](using c: chan.PCap[E, Branch[L, R]]^) |\label{line:chanbranch}|
      (l:  chan.PCap[E, L] ?=!> F)(r: chan.PCap[E, R] ?=!> F): F @kill(c) =
    if (...) then l(...) else r(...)  // invokes l or r depending on received choice
  // protocol recursion
  def recPush[E, T](): chan.PCap[E, Rec[T]] ?=!>? chan.PCap[T :: E, T] = ... |\label{line:chanrecpush}|
  def recTop[E, T](): chan.PCap[T :: E, Var[0]] ?=!>? chan.PCap[T :: E, T] = ... |\label{line:chanrectop}|
  def recPop[E, T, N <: Int](): chan.PCap[T :: E, Var[S[N]]] ?=!>? chan.PCap[E, Var[N]] = ...  |\label{line:chanrecpop}|
\end{lstlisting}
  
\end{subfigure}
\caption{Session Type and Chan API Definitions}
\label{fig:sessiondef}
\label{fig:sessionapi}
\end{figure}

We present an implementation of binary session types in our programming model, following 
\citet{DBLP:conf/icfp/JespersenML15, DBLP:conf/haskell/PucellaT08}, translated into Scala. Our definitions (\Cref{fig:sessiondef}) 
closely mirror the standard formulation: the @Local@ (line \ref{line:sessiontrait}) session type 
and extensions describe the protocol,
and the channel class @Chan@ (line \ref{line:chanclass}) embeds the protocol @T@ into a type member @PCap@, as capabilities.
Protocol recursion is represented by an 
additional \emph{protocol environment} parameter @E@ (line \ref{line:chanpcap}), 
which stores protocols later retrievable by de-Bruijn indices. 
The type @Dual@, recursively computing the duality of a given @Local@ type, 
is implemented using Scala's \emph{match types} \cite{DBLP:journals/pacmpl/BlanvillainBKO22}.

\Cref{fig:sessionapi} then defines the @Chan@ API.
The factory method for channels (line \ref{line:chanapply}) returns a pair of 
@Sigma@ objects, each containing a @Chan@ object in conjunction with its @PCap@ capability. The two 
endpoints have dual protocols, and both @PCap@ capabilities become available as implicits in the caller's scope. 
The @send@ method (line \ref{line:chansend}) requires a @Send[B, T]@-typed capability, and transitions the protocol 
to only @T@. Because @recv@ (line \ref{line:chanrecv}) must also return a value of type @B@,
we cannot use @?=!>?@ directly. Instead, @?=!>@ and @?<=@ are used to signify that it 
takes in a @Recv[B, T]@-typed capability, revokes it, and then returns the 
remaining session capability implicitly and the received value explicitly. 

To handle branching, we must return either an @L@-typed capability or an @R@-typed capability.
Therefore, @branch@ takes two callbacks and invokes one depending on the received choice.
To work with Scala's type inference, @branch@ must specify an explicit @using@ clause and a kill effect.

The methods @recPush@ (line \ref{line:chanrecpush}), @recTop@ (line \ref{line:chanrectop}), and 
@recPop@ (line \ref{line:chanrecpop}) carry out protocol recursion. First, @recPush@ operates on a @Rec[T]@-typed @Chan@
by pushing @T@ onto the protocol environment @E@. Then, a @Var[K]@-typed @Chan@ refers to the 
protocol @K@ deep in @E@. If @K@ is @0@, then @recTop@ replaces @Var[0]@ with the top of @E@. 
Otherwise, @K@ could be the successor of a natural number @N@, expressed in Scala as @Var[S[N]]@. Here,
@recPop@ can be used to remove the top of @E@ and decrement @N@.

\begin{figure}[b]
\begin{subfigure}[t]{0.45\linewidth}
\begin{lstlisting}[style=impl,numbers=none]
def echoServer(chan: Chan): ... = 
  chan.recPush()
  def recur(chan: Chan): ... = 
    println(chan.recv())
    chan.branch {
      chan.recTop(); recur(chan)
    } {
      chan.close()
    }
  recur(chan)
\end{lstlisting}
\vspace{-1ex}
\caption{Echo Server implementation. It prints out the received message 
and then either repeats or closes depending on client choice.}
\end{subfigure} \hspace{0.7cm}
\begin{subfigure}[t]{0.45\linewidth}
\begin{lstlisting}[style=impl,numbers=none]
def echoClient(chan: Chan): ... =
  chan.recPush()
  def recur(chan: Chan): ... =
    chan.send(readLine()) 
    if (readLine() == ...) then
      chan.left(); chan.recTop()
      recur(chan)
    else
      chan.right(); chan.close()
  recur(chan)
\end{lstlisting}
\vspace{-1ex}
\caption{Echo Client implementation. It reads and sends a message
before repeating or closing depending on client choice}
\label{fig:echoclient}
\end{subfigure}
\begin{subfigure}[b]{\linewidth}
\medskip
\begin{lstlisting}[style=impl,numbers=none]
def main() =
  val (serverChan, clientChan) = Chan[EchoServer]() 
  cFuture { echoServer(serverChan) }  // kills serverChan.PCap, a free variable of body
  cFuture { echoClient(clientChan) }  // kills clientChan.PCap
\end{lstlisting}
\vspace{-1ex}
\caption{Running channels in parallel. Using \lstinline|cFuture| permits typestate transitions.}
\label{fig:runchannels}
  
\end{subfigure}
\caption{Echo program implementation. The method return types are omitted.}
\label{fig:echoprogram}
\end{figure}

\paragraph{Echo Server/Client.}
We demonstrate this API in \Cref{fig:echoprogram}.
The server will receive a string, print it and then, depending on 
client choice, either repeat or quit. This protocol can be defined as:
\begin{lstlisting}
type EchoSInner = Recv[String, Branch[Var[0], End]]
type EchoServer = Rec[EchoSInner]; type EchoClient = Dual[EchoServer]
\end{lstlisting}
The methods @echoServer@ and @echoClient@ require a @PCap@ capability 
for their respective protocols, and to perform state transitions 
on the channel, they must also kill it, resulting in a full signature of:
\begin{lstlisting}
def echoServer(chan: Chan): chan.PCap[PNil, EchoServer] ?=!> Unit
\end{lstlisting}
The inner @recur@ methods also have a similar return type signature.
Then, we can create a channel and run @echoServer@ and @echoClient@ in parallel (\Cref{fig:runchannels}).
The method @cFuture@ is a wrapper for creating @Future@. 
As callbacks passed to @cFuture@ may 
kill a free variable (the captured @PCap@ capabilities), @cFuture@ 
possesses an observable kill effect that must be annotated
on the callback type. We introduce a \emph{function self-reference} @FUN@ for \lstinline|@kill()|;
a function with \lstinline|@kill(FUN)| will induce a destructive effect on itself,
thus allowing it to kill arbitrary free variables. The method @cFuture@ is then defined as: 
\begin{lstlisting}
def cFuture[T](body: => T @kill(FUN)): Future[T]  // body can only be used once
\end{lstlisting}
where the @FUN@ marker refers to @body@ itself. Note that @body@ is a by-name parameter:
any expression provided to @cFuture@ as a parameter will be lifted into
a parameterless function, subject to destruction by \lstinline|@kill(FUN)|.
Used in \Cref{fig:runchannels}, this ensures that the server and the client are
both one-shot.

\subsection{Control Flow} \label{sec:controlflow}
An important facet of typestate is its interaction with control flow.  
While our model supports basic typestate use within conditionals 
(see \Cref{fig:domexample,fig:echoprogram}), 
some usages are not directly expressible. We address this by defining 
a series of generic combinators for control flow (\Cref{fig:controlflowdefs}).  

\begin{figure}[t]
\centering
\begin{lstlisting}[style=impl]
def ifDiff[T, B1, B2](using c: T^)(cond: => Boolean)[U] |\label{line:ifDiff}|
  (tbranch: (T^) ?=!> ((B1^) ?<= U))
  (ebranch: (T^) ?=!> ((B2^) ?<= U)): ((Either[B1, B2]^) ?<= U) @kill(c) = ...
def matchSame[B1, B2, T](using c: Either[B1, B2]^)[U] |\label{line:matchSame}|
  (left:  (B1^) ?=!> ((T^) ?<= U))
  (right: (B2^) ?=!> ((T^) ?<= U)): (T^ ?<= U) @kill(c) = ...
def loop[T](using c: T^)(cond: => Boolean)(body: T ?=!>? T):  ((T^) ?<= Unit) @kill(c) = ... |\label{line:loop}|
def whileLeft[T, U](using c: T^)(body: T ?=!>? Either[T, U]): ((U^) ?<= Unit) @kill(c) = ... |\label{line:whileLeft}|
\end{lstlisting}
\caption{Control Flow Combinators}
\label{fig:controlflowdefs}
\end{figure}

The method @ifDiff@ (line \ref{line:ifDiff}) is parameterized by a capability of
type @T@, meant to represent some arbitrary initial typestate. It then evaluates
either @tbranch@ or @ebranch@ depending on the result of @cond@. Both thunks can
transition the initial typestate @T@ to some different states, @B1@ for
@tbranch@ and @B2@ for @ebranch@. Then, the union of the two typestates is
represented by implicitly returning an @Either[B1,B2]@. The method @matchSame@
(line \ref{line:matchSame}) is the dual of @ifDiff@.  It performs the inverse
operation of deconstructing an @Either@, requiring both cases to transition to
the same typestate.

Furthermore, @ifDiff@ and @matchSame@ can be used to define methods for
iteration. The method @loop@ (line \ref{line:loop}) takes in some state @T@, and
then repeats @body@ while @cond@ remains true. Importantly, @body@ must maintain
the state @T@ to ensure sound repetition. The method @whileLeft@ (line
\ref{line:whileLeft}) has no condition; it repeats @body@ until the initial
state @T@ transitions to some other state @U@.
The implementation of these combinators \cite{artifact} illustrates
an advantage of using capabilities: since they are merely program values,
they can be easily inspected and manipulated by
ordinary constructs, \eg, pattern matches.  

\begin{figure}[h]
\centering
\begin{subfigure}{\textwidth}
\centering
\begin{subfigure}{0.40\textwidth}
\begin{lstlisting}[style=impl,numbers=none]
if (...) { f.open() }  // f: File
   else  { f.open(); f.close() } 
\end{lstlisting}
\end{subfigure}
\hspace{1em}
\begin{subfigure}{0.50\textwidth}
\begin{lstlisting}[style=impl,numbers=none]
ifDiff[f.IsClosed, f.IsOpen, f.IsClosed] (...)
  { f.open() } { f.open(); f.close() }
\end{lstlisting}
\end{subfigure}
\vspace{-1ex}
\caption{Joining two different typestates. Using \lstinline|ifDiff| 
returns an \lstinline|Either[...]| for later use.}
\label{fig:join_ts}
\smallskip
\end{subfigure}

\begin{subfigure}{\textwidth}
\centering
\begin{subfigure}{0.40\textwidth}
\begin{lstlisting}[style=impl,numbers=none]
if (...) { f.open(); 10 } 
    else { f.open(); 20 } 
\end{lstlisting}
\end{subfigure}
\hspace{1em}
\begin{subfigure}{0.50\textwidth}
\begin{lstlisting}[style=impl,numbers=none]
ifDiff[f.IsClosed, f.IsOpen, f.IsOpen] (...)
  { f.open(); 10 } { f.open(); 20 }
\end{lstlisting}
\end{subfigure}
\vspace{-1ex}
\caption{While returning other data, using \lstinline|ifDiff|
recovers state transition via \lstinline|Sigma|-lifting.}
\label{fig:retrieve_ts}
\smallskip
\end{subfigure}

\begin{subfigure}{\linewidth}
\centering
\begin{subfigure}{0.40\textwidth}
\begin{lstlisting}[style=impl,numbers=none]
// next() transitions it.HasMore  
//                 to it.HasMore
loop[it.HasMore] (it.hasNext()) {
  val item = it.next()
  /* further computation... */ }
\end{lstlisting}
\end{subfigure}
\hspace{1em}
\begin{subfigure}{0.50\textwidth}
\begin{lstlisting}[style=impl,numbers=none]
// nextChecked() transitions it.HasMore  
//                 to Either[it.HasMore, it.End]
whileLeft[it.HasMore, it.End] {
  val item = it.nextChecked()
  /* further computation... */ }
\end{lstlisting}
\end{subfigure}
\vspace{-1ex}
\caption{Traversing through an iterator, \lstinline|it|, using either \lstinline|loop| or \lstinline|whileLeft|.}
\label{fig:iterationexample}
\end{subfigure}

\caption{Examples of using control flow combinators.}
\label{fig:ifdiffexample}
\vspace{-1ex}
\end{figure}
\Cref{fig:ifdiffexample} compares @ifDiff@ to @if@.
Without specialized type inference support,
we explicitly specify type arguments for @ifDiff@. 
The first argument is the initial state, the second the state after @tbranch@, and the third 
the state after @ebranch@. 
\Cref{fig:join_ts} demonstrates a conditional that closes @f: File@ in
one branch and opens it in another. A regular @if@ cannot express this,
as the return type would be a union of two different @Sigma@ types,
eventually @Any@, which is unusable.
In contrast, @ifDiff@ returns an @Either@ capability
that can later be consumed by @matchSame@. 

Alternatively, \Cref{fig:retrieve_ts} performs the same state transition
in both branches but returns an @Int@. For @if@, knowledge that a transition occurred is invisible
externally because the transition's resulting capability is local to each branch. However, @ifDiff@ can recover this information
by using @Sigma@ lifting to bundle the resulting capability with the returned value.  

\Cref{fig:iterationexample} demonstrates the iteration combinators @loop@ and 
@whileLeft@ by traversing through an iterator @it@ while tracking typestate.
Iterators have two states: @HasMore@ indicating more items and @End@ marking no more. 
Retrieving the next item can be done either via @next()@, unsafe and only transitioning from @HasMore@ to
@HasMore@, or with @nextChecked()@, transitioning to @HasMore@ or @End@. Then, @loop@ can 
be used with @next()@, as it requires the same iterator state. On the other hand, @whileLeft@ naturally
fits @nextChecked()@, repeating @nextChecked()@ until 
it transitions the state to @it.End@.
In both cases, @Sigma@-lifting ensures proper capabilities for further computation.

\pgfplotsset{compat=1.18}
\definecolor{pt-blue}{RGB}{68, 119, 170}
\definecolor{pt-cyan}{RGB}{102, 204, 238}
\definecolor{pt-green}{RGB}{34, 136, 51}
\definecolor{pt-purple}{RGB}{170, 51, 119}
\definecolor{pt-gray}{RGB}{187, 187, 187}

\section{Implementation and Discussion} \label{sec:formalimpl}

We have implemented a prototype in a fork of the Scala 3 compiler with the experimental capture checker enabled.
Our extensions to the capture checker are minor and non-invasive: the existing capture checking test suite (373 tests) continues to pass.
The implementation leverages the existing capture checker for capturing types \cite{DBLP:journals/toplas/BoruchGruszeckiOLLB23,DBLP:conf/scala/OderskyBBLL21,DBLP:journals/pacmpl/BrachthauserSLB22} to realize a simplified fragment of reachability types \cite{DBLP:journals/corr/abs-2510-08939}, 
where the distinctions are immaterial.
In this section, we discuss the intended safety properties of our approach,
along with its theoretical and practical limitations.

\subsection{Intended Safety Properties} \label{sec:safety-properties}

As this work primarily focuses on language design, a complete formalization integrating reachability types, path-dependent types, implicit resolution, and ANF transformation is beyond the scope of this paper.
Nevertheless, we can characterize the intended safety properties by examining the three core capability operations in our framework: \emph{receive}, \emph{revoke}, and \emph{return}.
\begin{itemize}[leftmargin=10pt]
  \item \textbf{Receive}: a capability can only be received while it is live, \ie, associated with its object via path-dependent type and not yet revoked.
  \item \textbf{Revoke}: once revoked, a capability is permanently unavailable. Following the rules of destructive effects, a revoked capability cannot be used explicitly or summoned via implicit resolution.
  \item \textbf{Return}: a returned capability is inserted into the implicit scope at the call site, then the compiler can automatically supply it to any subsequent operations that require it until revoked.
\end{itemize}

Although a full soundness proof is beyond the scope of this paper,
we empirically validate our approach through the case studies in \Cref{sec:casestudy}.
We refer interested readers to our implementation and accompanying examples
for details \cite{artifact}.

\subsection{Design and Limitations}

Our prototype introduces two main extensions to the compiler.
The \emph{destructive effect checker} is implemented as a compiler phase directly after capture checking,
architected as a bidirectional typer \cite{DBLP:journals/csur/DunfieldK21,DBLP:conf/popl/OderskyZZ01}.
It re-types the capture-checked syntax tree while recording a set of killed capabilities, promulgating kill annotations to types.
The \emph{type-directed ANF transform} \cite{DBLP:conf/pldi/FlanaganSDF93,DBLP:conf/icfp/RompfMO09} is implemented in the Scala typer,
triggered when encountering expressions of type @Sigma@.
As the transformation lifts such expressions, preservation of evaluation order is ensured by marking the generated bindings as @lazy@.
We discuss the design choices and limitations below.

\paragraph{Simplification on Destructive Effects.}\label{subsec:destructive-effect}
Our destructive effect checker draws on \citet{DBLP:journals/corr/abs-2510-08939},
which formalizes \emph{use} and \emph{kill} effects on top of System \Fsub with higher-order references and reachability qualifiers.
Our implementation adopts a simplified subset sufficient for modeling capability revocation, which introduces several limitations.

First, we omit the explicit \emph{use} effect and instead conservatively approximate usage via the full \emph{reachability} mentioned in qualifiers. 
While \citet{DBLP:journals/corr/abs-2510-08939} distinguishes between mentioning and actual use, allowing only uses of killed resources to be forbidden, 
the precision requires latent effect annotations on function types and
significantly increases the complexity of the type system,
which is unnecessary for our minimalistic extension.

Second, we omit mutable state, thus forbidding destructive effects on mutable variables and object fields.
However, objects with tracked typestate can still be stored in mutable variables and undergo state transitions, since revocation targets
their capabilities but not themselves.
Such variables must be ascribed to corresponding singleton types to
preserve the link to their capabilities.

Third, \citet{DBLP:journals/corr/abs-2510-08939} supports \emph{one-shot functions} that consume/kill captured free variables by using
explicitly named self-references in latent effects.
Our implementation adopts a simplified static notation @FUN@ to denote the self-reference at the innermost (most recent) level.
This is limited but practical enough for our implementation atop the capture checker,
and can be improved by building on a full reachability type checker.

\paragraph{Sigma.}
The @Sigma@ construct serves as the device for returning bundled capabilities.
However, it is not directly expressible using reachability types, which lack dependent types and cannot yet track fresh identities within data structures.
As a workaround, functions returning @Sigma@ should be understood in continuation-passing style, providing capabilities as fresh to the continuation.
Accordingly, @Sigma@ should be viewed as a transient wrapper that requires immediate unpacking after return:
defining a method that returns @Sigma@ requires providing a capability, while calling such a method introduces the implicit capability at the call site.

\paragraph{Error Reporting.}
Our approach encodes typestate using existing language features rather than introducing it as a built-in primitive.
As a result, error messages reflect the underlying encoding, which requires some familiarity to interpret,
a tradeoff commonly seen in encoding-based approaches \cite{DBLP:conf/icfp/JespersenML15, DBLP:conf/ecoop/ScalasY16}.
For instance, in our encoding of session types, performing the same action twice on a channel reports the use of a killed variable, 
which is meaningful once one understands that the first action kills the channel's capability.

More specifically, typestate violations usually reduce to two kinds of errors, effect-system errors, or implicit-search failures. %
For example:
\begin{lstlisting}
def g(f: File): f.IsOpen^ ?=> Unit =
  f.close()  // Error as f performs state transition
             //   not reflected in method signature
${\color{red} \text{Type Mismatch Error:}}$ Found: (c: f.IsOpen^) ?=> Unit @kill(c); Expected: f.IsOpen^ ?=> Unit
\end{lstlisting}
Here, a typestate transition induces a kill effect not reflected in the method signature, which is then surfaced as a type mismatch at the source position of the body of @g@.
Likewise:
\begin{lstlisting}
withFile { (f: File) => (c: f.IsClosed^) =>
  ()  // no implicit f.IsClosed capability for Sigma-lifting
}
${\color{red}\text{Sigma-lifting failed:}}$ No ${\color{black} \text{given instance of type}}$ f.IsClosed^ was found.
\end{lstlisting}
This example yields an implicit-search failure at the source position of @()@.

\begin{figure}[t]%
    \centering
    \begin{tikzpicture}
    \begin{axis}[
        ybar stacked,
        width=0.5\textwidth,      %
        height=7cm,                %
        bar width=14pt,            %
        enlarge x limits={abs=0.5cm}, 
        ylabel={Average Time (ms)},
        ylabel style={font=\footnotesize},
        y tick label style={
            font=\footnotesize,
            /pgf/number format/1000 sep={}
        },
        symbolic x coords={D16E, D16I, D16F, D24F, D32F},
        xtick=data,
        xticklabels={17E, 17I, 17F, 25F, 33F},
        x tick label style={font=\scriptsize},
        ymajorgrids=true,
        legend style={
            at={(0.03,0.97)},        
            anchor=north west,      
            legend columns=1,       
            font=\footnotesize,            %
            draw=none,              
            fill=white,             
            fill opacity=0.8,       
            text opacity=1,
            cells={anchor=west}
        },
        ymin=0,
        error bars/error bar style={color=black, thin},
        error bars/error mark options={color=black, thin, mark size=1.5pt}
    ]
    
    \addplot[fill=pt-blue, draw=none, error bars/.cd, y dir=both, y explicit] coordinates {
        (D16E, 275.20) +- (0, 9.94)  
        (D16I, 499.78) +- (0, 5.16)   
        (D16F, 795.26) +- (0, 20.43)  
        (D24F, 866.98) +- (0, 31.79) 
        (D32F, 919.60) +- (0, 27.66)
    };
    
    \addplot[fill=pt-purple, draw=none, error bars/.cd, y dir=both, y explicit] coordinates {
        (D16E, 0.00)   +- (0, 0.00)  
        (D16I, 0.00)   +- (0, 0.00)   
        (D16F, 535.88) +- (0, 24.82)  
        (D24F, 806.25) +- (0, 34.29) 
        (D32F, 1174.76) +- (0, 25.38)
    };
    
    \addplot[fill=pt-green, draw=none, error bars/.cd, y dir=both, y explicit] coordinates {
        (D16E, 0.00)   +- (0, 0.00)  
        (D16I, 0.00)   +- (0, 0.00)   
        (D16F, 364.00) +- (0, 18.87)  
        (D24F, 832.79) +- (0, 36.62) 
        (D32F, 1795.88) +- (0, 100.62)
    };
    
    \addplot[fill=pt-cyan, draw=none, error bars/.cd, y dir=both, y explicit] coordinates {
        (D16E, 492.64) +- (0, 29.98) 
        (D16I, 558.73) +- (0, 19.88)  
        (D16F, 801.42) +- (0, 19.35)  
        (D24F, 928.27) +- (0, 24.43) 
        (D32F, 1102.06) +- (0, 24.52)
    };
    
    \legend{Typing, CC, Effects, Others}
    \end{axis}
    \end{tikzpicture}
    \caption{Compilation time breakdown. Average of 10 runs, measured on macOS Sequoia 15.7.4 (Apple M3 Pro, 18 GB).
     A black tick above each compilation phase subcategory indicates standard deviation.}
    \label{fig:barchart}
\end{figure}

While error messages may grow more complex for sophisticated typestates,
users need not reason about low-level details such as the ANF transformation of @Sigma@,
since source-level metadata (\eg, source positions) can be propagated through @Sigma@ to present diagnostics at the appropriate level of abstraction.
Moreover, static typestate tracking shifts many errors to compile time, substantially reducing the need for runtime debugging.
We expect error reporting can be further improved with additional engineering effort. %

\paragraph{Performance Evaluation.}

We evaluate the performance of our compiler extensions with several benchmarks based on the @DOM@ tree API (cf. \Cref{fig:domdefsandapi})
as it is capable of tracking arbitrarily large typestate using type-level lists. \Cref{fig:barchart} 
breaks down the compilation time of our benchmarks by phase. All benchmarks are variations on benchmark \textbf{17F}, which
is a program that contains the code for \Cref{fig:domexample} and creates a @DOM@ tree of depth 17.
Benchmark \textbf{17I} removes
@Sigma@, capture checking, and effect checking, only
tracking implicit path-dependent capabilities, and benchmark \textbf{17E} further removes capabilities. Benchmarks \textbf{25F} 
and \textbf{33F} are based on \textbf{17F} but create a @DOM@ tree tracking 25 and 33 elements, respectively.
For medium-sized typestate (tracking 17 and 25 elements), the overhead introduced by our additions
(effect checking and the @Sigma@-directed ANF transform) is comparable to the overhead introduced by the capture checker, which is reasonable.
For larger typestate (\eg, \textbf{33F}), our effect-checking overhead is more visible than that of the capture checker, which can be mitigated by
further engineering effort.
\section{Related Work} \label{sec:related}

\paragraph{Representing Effects.}

A generic treatment of sequential (flow-sensitive) effect systems was explored
by \citet{DBLP:journals/toplas/Gordon21} using effect quantales, an algebraic
structure equipped with a sequencing operator for composing effects in order.
Earlier work in higher-order languages primarily addressed
flow-insensitive effects.
\citet{DBLP:conf/popl/LucassenG88} proposed a polymorphic calculus with effects
regarding region-based memory management. Generalized from regions,
\citet{hengleinEffectTypesRegionBased2005} introduced a calculus that represents
effects with scope tags. \citet{DBLP:conf/tldi/MarinoM09} characterized
effects using \emph{check} and \emph{adjust}
to manage capabilities.  \citet{DBLP:conf/popl/LindleyMM17,
DBLP:journals/pacmpl/BrachthauserSLB22,DBLP:journals/pacmpl/TangWDHLL25} 
further use capabilities or ambient effects to avoid
effect polymorphism.

Within the Scala ecosystem, \citet{DBLP:conf/ecoop/RytzOH12} introduced relative
effect polymorphism to alleviate the annotation overhead for effects.
Subsequently, \citet{DBLP:conf/oopsla/ToroT15} proposed a gradual
effect system that integrates static and dynamic effect checking. More recently,
effect handlers have been realized as a Scala
library~\cite{DBLP:journals/pacmpl/BrachthauserSO18,DBLP:journals/jfp/BrachthauserSO20},
leveraging capabilities to manage and control effects.

Continuation-passing style (CPS) transformations~%
\cite{DBLP:journals/mscs/DanvyF92} and
monads~\cite{DBLP:conf/popl/Filinski99,DBLP:conf/popl/Filinski94} have
well-established connections to effects~\cite{DBLP:conf/lfp/DanvyF90,
DBLP:conf/popl/Wadler92,DBLP:journals/tocl/WadlerT03}.
Of particular relevance, \citet{DBLP:conf/icfp/RompfMO09} introduced a
selective CPS transformation to implement a polymorphic calculus
with @shift/reset@ \cite{DBLP:conf/lfp/DanvyF90}, based on a
flow-sensitive effect system inspired by earlier work \cite{DBLP:conf/aplas/AsaiK07}.
In contrast, composing monads is
known to be difficult, requiring additional mechanisms
\cite{DBLP:conf/popl/LiangHJ95,DBLP:journals/jfp/Swierstra08,
DBLP:conf/haskell/KiselyovI15}.

In this work, we model effects through capabilities. The introduction of
capabilities is enabled by a selective ANF transformation, while their
revocation is managed by a destructive effect system.

\vspace{-2pt}
\paragraph{Tracking Typestate.}
\citet{DBLP:journals/tse/StromY86} introduced the concept of typestate,
initially assuming a setting where aliasing could be statically resolved,
which is generally not feasible in the presence of pointers.
To address typestate in the presence of aliasing, subsequent work has
employed whole-program analyses~\cite{DBLP:conf/oopsla/NaeemL08,
DBLP:conf/ppdp/JakobsenRD21,DBLP:journals/tosem/FinkYDRG08}.
\citet{DBLP:conf/ecoop/DeLineF04} proposed modular typestate checking by
distinguishing non-aliased objects, thereby enabling typestate
enforcement in those cases.

For aliased objects, \citet{DBLP:conf/oopsla/BierhoffA07} advanced the use of
fractional capabilities in linear reasoning, which underpins
the typestate-oriented programming language,
Plaid~\cite{DBLP:conf/oopsla/AldrichSSS09, DBLP:journals/toplas/GarciaTWA14}.
In addition to state transitions, Plaid requires annotating access permissions
of arguments to support modular aliasing reasoning; our approach does not
require such annotations.
\citet{DBLP:journals/pacmpl/Saffrich0024} further proposed replacing
transition annotations with ordered handling of borrows.

Session types~%
\cite{DBLP:conf/concur/Honda93,DBLP:conf/parle/TakeuchiHK94,
DBLP:conf/esop/HondaVK98} exemplify typestate reasoning by enforcing type-safe
communication protocols between concurrent processes. Beyond communication,
\citet{DBLP:conf/popl/GayVRGC10} extended session types to specify object
protocols, thereby achieving expressiveness comparable to general typestate
systems, with restrictions on object aliasing.
While session types usually require passing channels for
linear reasoning, proposals have been made \cite{DBLP:conf/ppdp/Saffrich023,
DBLP:journals/lmcs/SaffrichT22,DBLP:journals/tcs/VasconcelosGR06}
to enable direct-style programming.

Session types have been implemented
in several languages, such as Rust \cite{DBLP:conf/icfp/JespersenML15,DBLP:journals/corr/abs-1909-05970},
Haskell \cite{DBLP:conf/haskell/PucellaT08}, and Scala \cite{DBLP:conf/ecoop/ScalasY16};
our example is based on the Rust and Haskell implementations. The Scala
implementation \cite{DBLP:conf/ecoop/ScalasY16} relies on run-time
enforcement of channel linearity.

\vspace{-2pt}
\paragraph{Linear Types and Fractional Capabilities.}
Linear logic~\cite{DBLP:journals/tcs/Girard87} restricts the structural rules of
contraction and weakening, thereby controlling the duplication and disposal of
resources. This logical foundation underpins linear type
systems~\cite{DBLP:conf/ifip2/Wadler90}, in which values must be used exactly once,
facilitating safe and predictable management of side
effects and mutable state.

Although resources in practical programming are often shared and thus nonlinear,
their capabilities~\cite{DBLP:journals/cacm/DennisH66,
DBLP:conf/asian/MillerS03} can be abstracted and managed linearly.
Such capabilities can be made fractional \cite{DBLP:conf/sas/Boyland03}
through splitting and rejoining.
For example, the type system of Rust~\cite{DBLP:conf/sigada/MatsakisK14} permits
concurrent reads when write access is disabled;
full control is restored once all read borrows have ended.

The integration of substructural reasoning and fractional capabilities is common
in typestate analysis and session type systems for regulating shared access.
This approach manifests as access permissions in
Plaid~\cite{DBLP:journals/toplas/GarciaTWA14}, linear constraints in Linear
Haskell~\cite{DBLP:journals/pacmpl/SpiwackKBWE22}, and borrows from ordered
partial monoids~\cite{DBLP:journals/pacmpl/Saffrich0024,DBLP:journals/pacmpl/SaffrichS0V25}.
In this work, we employ capabilities without substructural
reasoning, and therefore do not impose multiplicity or ordering
constraints on their use.

\vspace{-2pt}
\paragraph{Scala.}
Scala is a programming language that integrates object-oriented and functional
paradigms, featuring an advanced static type system. Its type system is
formalized as the Dependent Object
Types (DOT) calculus~\cite{DBLP:conf/oopsla/RompfA16,DBLP:conf/birthday/AminGORS16}, 
which enables types to be parameterized by object paths, thereby supporting precise
path-dependent reasoning.

Scala additionally provides implicit argument
resolution~\cite{DBLP:journals/pacmpl/OderskyBLBMS18}, enabling the
automatic inference of function parameters based on type information. This
feature facilitates capability-based
programming~\cite{DBLP:conf/scala/OderskyBBLL21} by obviating the need for
explicit capability passing. To ensure that capabilities do not escape their
intended scopes, proposals have been made to track them as \emph{second-class}
values~\cite{DBLP:conf/oopsla/OsvaldEWAR16,DBLP:conf/ecoop/XhebrajB0R22} with
restricted usages and lifetimes. More recently, the Scala compiler
introduced an experimental capture checker~\cite{ScalaCC}, which employs
descriptive alias tracking to achieve similar safety guarantees.

\vspace{-2pt}
\paragraph{Descriptive Alias Tracking with Reachability/Capturing Types.}

Reachability and capturing type systems both aim to track resources in types, 
using sets of variable names as qualifiers.
Motivated by different use cases,
these systems primarily diverge in their treatment of unnamed
resources. Early proposals for capturing
types~\cite{DBLP:journals/toplas/BoruchGruszeckiOLLB23,DBLP:conf/scala/OderskyBBLL21,DBLP:journals/pacmpl/BrachthauserSLB22}
focused on preventing unintended escape of critical resources, notably
capabilities.
Polymorphism is ergonomically supported by boxing enclosed resources;
unboxing outside the intended scope is disallowed.
In contrast, reachability types~\cite{DBLP:journals/pacmpl/BaoWBJHR21}
track escaping
resources via self-references. Their original formulation omitted
polymorphism, which was subsequently addressed through
lightweight polymorphism and explicit
quantification~\cite{DBLP:journals/pacmpl/WeiBJBR24}.

Both reachability and capturing types continue to evolve,
concerning tracking separation \cite{DBLP:journals/pacmpl/WeiBJBR24,DBLP:journals/pacmpl/XuBO24},
mechanisms for implementation \cite{jia26escape,DBLP:journals/corr/abs-2306-06496},
and semantic denotation \cite{DBLP:journals/pacmpl/XuBPO25,DBLP:journals/pacmpl/BaoJ0BR25}.
The experimental Scala capture checker~\cite{ScalaCC} has evolved along several
lines of research and remains under active development; it has not yet
been described end-to-end in the literature.
For the purpose of this work, we focus on a common core
whose behavior aligns with reachability types.

Within the framework of reachability types, effect systems relative to
reachability-tracked values were informally envisioned from the outset,
with potential applications including Rust-style ownership transfer and move
semantics~\cite{DBLP:journals/pacmpl/BaoWBJHR21,DBLP:journals/pacmpl/WeiBJBR24}.
However, until recently, reachability-sensitive effect systems had been
neither fully formalized nor implemented in a widely-used language.
\citet{DBLP:journals/pacmpl/BracevacWJAJBR23} investigate compiler optimizations using an integrated
type-effect-dependency system. 
While \citet{he26arena} address
memory management via flow-insensitive scoped allocations with guaranteed
lexical deallocation, \citet{DBLP:journals/pacmpl/BaoJ0BR25} proved
effect safety for a flow-insensitive effect system using logical relations,
but without considering deallocation or other destructive effects.
Most relevant to our work, \citet{DBLP:journals/corr/abs-2510-08939} formalize 
flow-sensitive kill effects with sound deallocation, which serves as 
the foundation for our approach to typestate tracking via revocable 
capabilities. With the goal of putting theory into practice, 
and of studying the practical viability of the approach, the present 
paper supplies a prototype implementation
as an extension of the Scala 3 compiler.
\section{Conclusion} \label{sec:conc}

In this paper, we show that expressive, flow-sensitive typestate tracking is
possible with minimal extensions to existing capability-based systems.
By decoupling capability lifetimes from lexical scopes and supporting 
the revocation and implicit returning of capabilities,
our approach enables precise and safe management of
stateful resources in imperative code.
As key supporting mechanisms, our additions include a destructive effect system
and a type-directed ANF transformation.
The resulting Scala 3 prototype supports a variety of stateful patterns,
while maintaining concise and readable code.
This work bridges the gap between scoped reasoning and flow-sensitive
expressiveness, advancing the safety and ergonomics of stateful programming. %

\section*{Data Availability Statement}

Details of this work can be found in our artifact \cite{artifact}.
Our artifact is also available at \url{https://github.com/TiarkRompf/scala3/tree/artifact}.

\begin{acks}                            %
We thank anonymous reviewers for their valuable feedback.
This work was supported in part by \grantsponsor{NSF}{NSF}{} award
\grantnum{NSF}{2348334} and an Augusta University faculty startup package,
as well as gifts from Meta, Google, Microsoft, and VMware.
\end{acks}

\appendix
\clearpage
\section{Case Study Implementation Details} \label{appendix:case-study}
The low-level implementation details were omitted from the definitions in \Cref{sec:casestudy}.
In particular, we omit how to construct path-dependent capabilities and
how to perform typestate transitions, which we elaborate in this section. We only discuss
\Cref{sec:lockagain} and \Cref{sec:controlflow}, and other examples
share the same techniques.

\subsection{Table Locking}
\begin{figure}[h]
\begin{lstlisting}[style=impl]
trait Lock: 
  type IsHeld // lock is held, usable 
  type IsReleased // lock is released, unusable

class Table private[...](n: Int) extends Lock: 
  private val data = ... // an indexable data structure
  // ... table lock field ...
  class Row private[...](m: Int) extends Lock: 
    private val row = data(m) // mth row of table
    // ... row lock field ...   

object Table: 
  def apply(n: Int): Sigma { type A = Table; type B = a.IsReleased^ } = // factory method |\label{line:apptableapply}|
    val table = new Table(n) { type IsReleased = Unit; type IsHeld = Unit }  
    new Sigma {
      type A = Table; type B = a.IsReleased^ 
      val a: table.type = table; val b: a.IsReleased^ = () // Opaque outside of apply
    } 

extension (table: Table) 
  def lock(): table.IsReleased ?=!>? table.IsHeld = |\label{line:apptablelock}|
    // ... acquire the lock for table ... 
    Sigma((), ().asInstanceOf[table.isHeld^])  // safe type cast |\label{line:appsafecast}|

  def unlock(): table.IsHeld ?=!>? table.IsReleased =   // ... release the lock ... 

  def locateRow(n: Int): table.IsHeld^ ?=> Sigma { type A = table.Row; type B = a.IsReleased^ } = 
    val row = new table.Row(n) { type IsReleased = Unit; type IsHeld = Unit } 
    new Sigma { 
      type A = table.Row; type B = a.IsReleased^
      val a: row.type = row; val b: a.IsReleased^ = () }

  def lockRow(row: table.Row) table.isHeld^ ?=> row.IsReleased ?=!>? row.IsHeld = 
    Sigma((), ().asInstanceOf[row.IsHeld^])

extension (row: Table#Row) 
  def unlock(): row.IsHeld ?=!>? row.IsReleased = 
    Sigma((), ().asInstanceOf[row.IsReleased^])

  def computeOnRow(): row.IsHeld^ ?=> ... = ... // requires row capability 
\end{lstlisting}
\caption{Table locking definitions with further details}
\label{fig:tablelockdet}
\end{figure}

\Cref{fig:tablelockdet} repeats the table locking definitions with some implementation details filled in.  
To construct path-dependent capabilities, the factory method @apply@ (line \ref{line:apptableapply}) initializes
the type members @IsHeld@ and @IsReleased@ to @Unit@. Within @apply@, @()@ can be used as these capabilities,
so @apply@ returns a @Sigma@ with @a.IsReleased@ set to @()@. Importantly, this initialization is opaque outside of @apply@,
or in other words it is only known within @apply@. Another key point is that @Sigma@ only triggers a transform if it
is in non-tail position, so the @new Sigma@ here does not transform.

Methods that perform state transitions, such as @lock@ (line \ref{line:apptablelock}) must construct 
a new @Sigma@ explicitly returning @Unit@ and implicitly returning the path-dependent capability corresponding 
to the new state. Instead of directly invoking the constructor of @Sigma@, which involves manually 
initializing type members, we provide a method @Sigma@ defined as:
\begin{lstlisting}
  def Sigma[A1, B1](a1: A1, b1: B1): Sigma { type A = A1; type B = B1 } =
    new Sigma { type A = A1; type B = B1; val a: A1 = a1; val b: B1 = b1 }
\end{lstlisting} 

\noindent
To construct the capability corresponding to the new state, a type-cast (line \ref{line:appsafecast})
from @Unit@ to the capability is used. This is safe as the only way to introduce path-dependent capabilities
is the factory method @apply@, where the path-dependent capabilities are initialized to @Unit@. 

\subsection{Control Flow} \label{appendix:control-flow}
\begin{figure}[h]
\begin{lstlisting}[style=impl]
def move[T]: T ?=!>? T = // kills T and returns T fresh. |\label{line:appmove}|
  Sigma((), summon[T].asInstanceOf[T]) // safe type-cast

def ifDiff[T, B1, B2](using c: T^)(cond: => Boolean)[A1](tbranch: (T^) ?=!> ((B1^) ?<= A1)) |\label{line:appifdiff}|
    (ebranch: (T^) ?=!> ((B2^) ?<= A1)): ((Either[B1, B2]^) ?<= A1) @kill(c) =
  if cond then
    val a1 = tbranch(using c) // returns B1^ implicitly and A1 explicitly
    Sigma(a1, Left(summon[B1].asInstanceOf[B1]))
  else
    val a2 = ebranch(using c) // returns B1^ implicitly and A1 explicitly
    Sigma(a2, Right(summon[B2].asInstanceOf[B2]))

def matchSame[B1, B2, T](using c: Either[B1, B2]^)[U](left: (B1^) ?=!> ((T^) ?<= U)) |\label{line:matchSame}|
    (right: (B2^) ?=!> ((T^) ?<= U)): ((T^) ?<= U) @kill(c) =
  c match
    case Left(b1) => left(using b1.asInstanceOf[B1^]) 
    case Right(b2) => right(using b2.asInstanceOf[B2^])

def loop[T](using c: T^)(cond: => Boolean)(body: T ?=!>? T): ((T^) ?<= Unit) @kill(c) = |\label{line:apploop}|
  if cond then
    body(using c) 
    loop[T](cond)(body)
  else move[T] 
 
def whileLeft[T, U](using c: T^)(body: T ?=!>? Either[T, U]): ((U^) ?<= Unit) @kill(c) = |\label{line:appwhileLeft}|
  body(using c) 
  matchSame[T, U, U] { // (c: T^) ?=> 
    whileLeft[T, U](body) 
  } { move[U] } // (c: U^) ?=> move[U]
\end{lstlisting}
\caption{Control flow definitions with implementation.}
\label{fig:controlflowimpl}
\end{figure}

\Cref{fig:controlflowimpl} presents the implementation for 
several control flow combinators. Prior 
to defining combinators, it is helpful to first define a @move@ operation (line \ref{line:appmove})
that disables all accesses to some capability @T@ and returns it as contextually fresh.
With no actual move effects in Scala, we use a type cast to make the same capability appear
fresh for returning. The cast is safe because all previous aliases are killed.
Also note that @summon[T]@ suffices to summon a capability of type @T^@,
since implicit resolution occurs at the type-checking phase, where capture-checking syntax has no effect.

To implement @ifDiff@ (line \ref{line:appifdiff}), we use a regular conditional. It 
branches on @cond@, invokes the corresponding thunk, and then returns a corresponding @Sigma@. 
To work around the limitation of the capture checker, we use another type cast from @B1^@ to @B1@ for creating
a fresh @Either[...]^@. This is safe because the resulting @Either@ is still tracked.
As the dual of @ifDiff@, @matchSame@ works by matching on the @Either@ before invoking the thunks.
For the same reasons as @ifDiff@, a type cast is needed after deconstructing the @Either@.

The method @loop@ (line \ref{line:apploop}) also branches on @cond@. If @cond@ is true, then 
it calls @body@ before recursing. If it is false, then the capability that should be supplied to 
the result type of @T^ ?<= Unit@ is @c@. But @c@ cannot just be supplied, since otherwise 
the result type would not be @T^ ?<= Unit@, but rather @T^{c} ?<= Unit@. Therefore, we need to use @move@ 
to return something that is contextually fresh. This is safe because @c@ is killed by @loop@.  
Method @whileLeft@ is similar. It invokes @body@, and then uses @matchSame@ to match on the result.
If it is still the same capability @T@, then it recurses, and otherwise it returns @U@. 

As a larger example, we can rewrite the @main@ method (line: \ref{line:appdommain}) from the @DOM@ tree example using @loop@:
\begin{lstlisting}[style=impl]
// creates </tr><tr>
def nextTR[L <: EList](t: DOM): (t.Elems[TR :: L] ?=!>? t.Elems[TR :: L]) = 
  t.close(TR()); t.open(TR())

// creates <td>p._1</td> <td>p._2</td>
def twoCells[L <: EList](t: DOM, f: String, s: String): t.Elems[TR :: L] ?=!>? t.Elems[TR :: L] = 
  t.open(TD()); t.text(TD(), f); t.close(TD());
  t.open(TD()); t.text(TD(), s); t.close(TD()) 

def main() = makeDOM { dom =>  |\label{line:appdommain}|
  dom.open(TABLE()); dom.open(TBODY()); dom.open(TR()) 
  val response = ... ; val reader = ... ;

  type TStart = TR :: TBODY :: TABLE :: ENil
  var chunk = await(reader.read().toFuture)

  loop[dom.Elems[TStart]](!chunk.isDone) { // repeat body until chunk.isDone |\label{line:apploopdom}|
    val line = chunk.value; 
    twoCells(dom, line.timeStamp, line.message); 
    nextTR(dom);  // typestate stays fixed, e.g. dom.Elems[TStart] ?=!>? dom.Elems[TStart]
    chunk = await(reader.read().toFuture) // Sigma lifting will synthesize a dom.Elems[TStart]
  }

  dom.close(TR()); dom.close(TBODY()); dom.close(TABLE()) 
}
\end{lstlisting}

The relevant changes begin on line \ref{line:apploopdom}. Instead of 
defining a recursive function, we can use @loop@ to read and add each log file line 
to the body. We must explicitly provide the type argument to inform @loop@ what typestate 
should stay constant. Notably, the function @loop@ does not need to return a @Sigma@ directly,
since @Sigma@ lifting will automatically synthesize a @Sigma@ containing a @dom.Elems[TStart]@
capability.

\makeatother
\bibliography{references}


\begin{thebibliography}{77}


\ifx \showCODEN    \undefined \def \showCODEN     #1{\unskip}     \fi
\ifx \showISBNx    \undefined \def \showISBNx     #1{\unskip}     \fi
\ifx \showISBNxiii \undefined \def \showISBNxiii  #1{\unskip}     \fi
\ifx \showISSN     \undefined \def \showISSN      #1{\unskip}     \fi
\ifx \showLCCN     \undefined \def \showLCCN      #1{\unskip}     \fi
\ifx \shownote     \undefined \def \shownote      #1{#1}          \fi
\ifx \showarticletitle \undefined \def \showarticletitle #1{#1}   \fi
\ifx \showURL      \undefined \def \showURL       {\relax}        \fi
\providecommand\bibfield[2]{#2}
\providecommand\bibinfo[2]{#2}
\providecommand\natexlab[1]{#1}
\providecommand\showeprint[2][]{arXiv:#2}

\bibitem[Aldrich et~al\mbox{.}(2009)]%
        {DBLP:conf/oopsla/AldrichSSS09}
\bibfield{author}{\bibinfo{person}{Jonathan Aldrich}, \bibinfo{person}{Joshua
  Sunshine}, \bibinfo{person}{Darpan Saini}, {and} \bibinfo{person}{Zachary
  Sparks}.} \bibinfo{year}{2009}\natexlab{}.
\newblock \showarticletitle{Typestate-oriented programming}. In
  \bibinfo{booktitle}{\emph{{OOPSLA} Companion}}. \bibinfo{publisher}{{ACM}},
  \bibinfo{pages}{1015--1022}.
\newblock
\href{https://doi.org/10.1145/1639950.1640073}{doi:\nolinkurl{10.1145/1639950.1640073}}


\bibitem[Amin et~al\mbox{.}(2016)]%
        {DBLP:conf/birthday/AminGORS16}
\bibfield{author}{\bibinfo{person}{Nada Amin}, \bibinfo{person}{Samuel
  Gr{\"{u}}tter}, \bibinfo{person}{Martin Odersky}, \bibinfo{person}{Tiark
  Rompf}, {and} \bibinfo{person}{Sandro Stucki}.}
  \bibinfo{year}{2016}\natexlab{}.
\newblock \showarticletitle{The Essence of Dependent Object Types}. In
  \bibinfo{booktitle}{\emph{A List of Successes That Can Change the World}}
  \emph{(\bibinfo{series}{Lecture Notes in Computer Science},
  Vol.~\bibinfo{volume}{9600})}. \bibinfo{publisher}{Springer},
  \bibinfo{pages}{249--272}.
\newblock
\href{https://doi.org/10.1007/978-3-319-30936-1\_14}{doi:\nolinkurl{10.1007/978-3-319-30936-1\_14}}


\bibitem[Asai and Kameyama(2007)]%
        {DBLP:conf/aplas/AsaiK07}
\bibfield{author}{\bibinfo{person}{Kenichi Asai} {and}
  \bibinfo{person}{Yukiyoshi Kameyama}.} \bibinfo{year}{2007}\natexlab{}.
\newblock \showarticletitle{Polymorphic Delimited Continuations}. In
  \bibinfo{booktitle}{\emph{{APLAS}}} \emph{(\bibinfo{series}{Lecture Notes in
  Computer Science}, Vol.~\bibinfo{volume}{4807})}.
  \bibinfo{publisher}{Springer}, \bibinfo{pages}{239--254}.
\newblock
\href{https://doi.org/10.1007/978-3-540-76637-7\_16}{doi:\nolinkurl{10.1007/978-3-540-76637-7\_16}}


\bibitem[Bao et~al\mbox{.}(2025)]%
        {DBLP:journals/pacmpl/BaoJ0BR25}
\bibfield{author}{\bibinfo{person}{Yuyan Bao}, \bibinfo{person}{Songlin Jia},
  \bibinfo{person}{Guannan Wei}, \bibinfo{person}{Oliver Bracevac}, {and}
  \bibinfo{person}{Tiark Rompf}.} \bibinfo{year}{2025}\natexlab{}.
\newblock \showarticletitle{Modeling Reachability Types with Logical Relations:
  Semantic Type Soundness, Termination, Effect Safety, and Equational Theory}.
\newblock \bibinfo{journal}{\emph{Proc. {ACM} Program. Lang.}}
  \bibinfo{volume}{9}, \bibinfo{number}{{OOPSLA2}} (\bibinfo{year}{2025}),
  \bibinfo{pages}{1837--1864}.
\newblock
\href{https://doi.org/10.1145/3763116}{doi:\nolinkurl{10.1145/3763116}}


\bibitem[Bao et~al\mbox{.}(2021)]%
        {DBLP:journals/pacmpl/BaoWBJHR21}
\bibfield{author}{\bibinfo{person}{Yuyan Bao}, \bibinfo{person}{Guannan Wei},
  \bibinfo{person}{Oliver Bra\v{c}evac}, \bibinfo{person}{Yuxuan Jiang},
  \bibinfo{person}{Qiyang He}, {and} \bibinfo{person}{Tiark Rompf}.}
  \bibinfo{year}{2021}\natexlab{}.
\newblock \showarticletitle{Reachability types: tracking aliasing and
  separation in higher-order functional programs}.
\newblock \bibinfo{journal}{\emph{Proc. {ACM} Program. Lang.}}
  \bibinfo{volume}{5}, \bibinfo{number}{{OOPSLA}} (\bibinfo{year}{2021}),
  \bibinfo{pages}{1--32}.
\newblock
\href{https://doi.org/10.1145/3485516}{doi:\nolinkurl{10.1145/3485516}}


\bibitem[Bierhoff and Aldrich(2007)]%
        {DBLP:conf/oopsla/BierhoffA07}
\bibfield{author}{\bibinfo{person}{Kevin Bierhoff} {and}
  \bibinfo{person}{Jonathan Aldrich}.} \bibinfo{year}{2007}\natexlab{}.
\newblock \showarticletitle{Modular typestate checking of aliased objects}. In
  \bibinfo{booktitle}{\emph{{OOPSLA}}}. \bibinfo{publisher}{{ACM}},
  \bibinfo{pages}{301--320}.
\newblock
\href{https://doi.org/10.1145/1297027.1297050}{doi:\nolinkurl{10.1145/1297027.1297050}}


\bibitem[Blanvillain et~al\mbox{.}(2022)]%
        {DBLP:journals/pacmpl/BlanvillainBKO22}
\bibfield{author}{\bibinfo{person}{Olivier Blanvillain},
  \bibinfo{person}{Jonathan~Immanuel Brachth{\"{a}}user},
  \bibinfo{person}{Maxime Kjaer}, {and} \bibinfo{person}{Martin Odersky}.}
  \bibinfo{year}{2022}\natexlab{}.
\newblock \showarticletitle{Type-level programming with match types}.
\newblock \bibinfo{journal}{\emph{Proc. {ACM} Program. Lang.}}
  \bibinfo{volume}{6}, \bibinfo{number}{{POPL}} (\bibinfo{year}{2022}),
  \bibinfo{pages}{1--24}.
\newblock
\href{https://doi.org/10.1145/3498698}{doi:\nolinkurl{10.1145/3498698}}


\bibitem[Boruch{-}Gruszecki et~al\mbox{.}(2023)]%
        {DBLP:journals/toplas/BoruchGruszeckiOLLB23}
\bibfield{author}{\bibinfo{person}{Aleksander Boruch{-}Gruszecki},
  \bibinfo{person}{Martin Odersky}, \bibinfo{person}{Edward Lee},
  \bibinfo{person}{Ondrej Lhot{\'{a}}k}, {and}
  \bibinfo{person}{Jonathan~Immanuel Brachth{\"{a}}user}.}
  \bibinfo{year}{2023}\natexlab{}.
\newblock \showarticletitle{Capturing Types}.
\newblock \bibinfo{journal}{\emph{{ACM} Trans. Program. Lang. Syst.}}
  \bibinfo{volume}{45}, \bibinfo{number}{4} (\bibinfo{year}{2023}),
  \bibinfo{pages}{21:1--21:52}.
\newblock
\href{https://doi.org/10.1145/3618003}{doi:\nolinkurl{10.1145/3618003}}


\bibitem[Boyland(2003)]%
        {DBLP:conf/sas/Boyland03}
\bibfield{author}{\bibinfo{person}{John Boyland}.}
  \bibinfo{year}{2003}\natexlab{}.
\newblock \showarticletitle{Checking Interference with Fractional Permissions}.
  In \bibinfo{booktitle}{\emph{{SAS}}} \emph{(\bibinfo{series}{Lecture Notes in
  Computer Science})}. \bibinfo{publisher}{Springer}, \bibinfo{pages}{55--72}.
\newblock
\href{https://doi.org/10.1007/3-540-44898-5\_4}{doi:\nolinkurl{10.1007/3-540-44898-5\_4}}


\bibitem[Bracevac et~al\mbox{.}(2023)]%
        {DBLP:journals/pacmpl/BracevacWJAJBR23}
\bibfield{author}{\bibinfo{person}{Oliver Bracevac}, \bibinfo{person}{Guannan
  Wei}, \bibinfo{person}{Songlin Jia}, \bibinfo{person}{Supun Abeysinghe},
  \bibinfo{person}{Yuxuan Jiang}, \bibinfo{person}{Yuyan Bao}, {and}
  \bibinfo{person}{Tiark Rompf}.} \bibinfo{year}{2023}\natexlab{}.
\newblock \showarticletitle{Graph IRs for Impure Higher-Order Languages: Making
  Aggressive Optimizations Affordable with Precise Effect Dependencies}.
\newblock \bibinfo{journal}{\emph{Proc. {ACM} Program. Lang.}}
  \bibinfo{volume}{7}, \bibinfo{number}{{OOPSLA2}} (\bibinfo{year}{2023}),
  \bibinfo{pages}{400--430}.
\newblock
\href{https://doi.org/10.1145/3622813}{doi:\nolinkurl{10.1145/3622813}}


\bibitem[Brachth{\"{a}}user et~al\mbox{.}(2022)]%
        {DBLP:journals/pacmpl/BrachthauserSLB22}
\bibfield{author}{\bibinfo{person}{Jonathan~Immanuel Brachth{\"{a}}user},
  \bibinfo{person}{Philipp Schuster}, \bibinfo{person}{Edward Lee}, {and}
  \bibinfo{person}{Aleksander Boruch{-}Gruszecki}.}
  \bibinfo{year}{2022}\natexlab{}.
\newblock \showarticletitle{Effects, capabilities, and boxes: from scope-based
  reasoning to type-based reasoning and back}.
\newblock \bibinfo{journal}{\emph{Proc. {ACM} Program. Lang.}}
  \bibinfo{volume}{6}, \bibinfo{number}{{OOPSLA}} (\bibinfo{year}{2022}),
  \bibinfo{pages}{1--30}.
\newblock
\href{https://doi.org/10.1145/3527320}{doi:\nolinkurl{10.1145/3527320}}


\bibitem[Brachth{\"{a}}user et~al\mbox{.}(2018)]%
        {DBLP:journals/pacmpl/BrachthauserSO18}
\bibfield{author}{\bibinfo{person}{Jonathan~Immanuel Brachth{\"{a}}user},
  \bibinfo{person}{Philipp Schuster}, {and} \bibinfo{person}{Klaus Ostermann}.}
  \bibinfo{year}{2018}\natexlab{}.
\newblock \showarticletitle{Effect handlers for the masses}.
\newblock \bibinfo{journal}{\emph{Proc. {ACM} Program. Lang.}}
  \bibinfo{volume}{2}, \bibinfo{number}{{OOPSLA}} (\bibinfo{year}{2018}),
  \bibinfo{pages}{111:1--111:27}.
\newblock
\href{https://doi.org/10.1145/3276481}{doi:\nolinkurl{10.1145/3276481}}


\bibitem[Brachth{\"{a}}user et~al\mbox{.}(2020)]%
        {DBLP:journals/jfp/BrachthauserSO20}
\bibfield{author}{\bibinfo{person}{Jonathan~Immanuel Brachth{\"{a}}user},
  \bibinfo{person}{Philipp Schuster}, {and} \bibinfo{person}{Klaus Ostermann}.}
  \bibinfo{year}{2020}\natexlab{}.
\newblock \showarticletitle{Effekt: Capability-passing style for type- and
  effect-safe, extensible effect handlers in Scala}.
\newblock \bibinfo{journal}{\emph{J. Funct. Program.}}  \bibinfo{volume}{30}
  (\bibinfo{year}{2020}), \bibinfo{pages}{e8}.
\newblock
\href{https://doi.org/10.1017/S0956796820000027}{doi:\nolinkurl{10.1017/S0956796820000027}}


\bibitem[Danvy and Filinski(1990)]%
        {DBLP:conf/lfp/DanvyF90}
\bibfield{author}{\bibinfo{person}{Olivier Danvy} {and}
  \bibinfo{person}{Andrzej Filinski}.} \bibinfo{year}{1990}\natexlab{}.
\newblock \showarticletitle{Abstracting Control}. In
  \bibinfo{booktitle}{\emph{{LISP} and Functional Programming}}.
  \bibinfo{publisher}{{ACM}}, \bibinfo{pages}{151--160}.
\newblock
\href{https://doi.org/10.1145/91556.91622}{doi:\nolinkurl{10.1145/91556.91622}}


\bibitem[Danvy and Filinski(1992)]%
        {DBLP:journals/mscs/DanvyF92}
\bibfield{author}{\bibinfo{person}{Olivier Danvy} {and}
  \bibinfo{person}{Andrzej Filinski}.} \bibinfo{year}{1992}\natexlab{}.
\newblock \showarticletitle{Representing Control: {A} Study of the {CPS}
  Transformation}.
\newblock \bibinfo{journal}{\emph{Math. Struct. Comput. Sci.}}
  \bibinfo{volume}{2}, \bibinfo{number}{4} (\bibinfo{year}{1992}),
  \bibinfo{pages}{361--391}.
\newblock
\href{https://doi.org/10.1017/S0960129500001535}{doi:\nolinkurl{10.1017/S0960129500001535}}


\bibitem[DeLine and F{\"{a}}hndrich(2004)]%
        {DBLP:conf/ecoop/DeLineF04}
\bibfield{author}{\bibinfo{person}{Robert DeLine} {and} \bibinfo{person}{Manuel
  F{\"{a}}hndrich}.} \bibinfo{year}{2004}\natexlab{}.
\newblock \showarticletitle{Typestates for Objects}. In
  \bibinfo{booktitle}{\emph{{ECOOP}}} \emph{(\bibinfo{series}{Lecture Notes in
  Computer Science}, Vol.~\bibinfo{volume}{3086})}.
  \bibinfo{publisher}{Springer}, \bibinfo{pages}{465--490}.
\newblock
\href{https://doi.org/10.1007/978-3-540-24851-4\_21}{doi:\nolinkurl{10.1007/978-3-540-24851-4\_21}}


\bibitem[Deng et~al\mbox{.}(2025)]%
        {DBLP:journals/corr/abs-2510-08939}
\bibfield{author}{\bibinfo{person}{Haotian Deng}, \bibinfo{person}{Siyuan He},
  \bibinfo{person}{Songlin Jia}, \bibinfo{person}{Yuyan Bao}, {and}
  \bibinfo{person}{Tiark Rompf}.} \bibinfo{year}{2025}\natexlab{}.
\newblock \showarticletitle{Free to Move: Reachability Types with
  Flow-Sensitive Effects for Safe Deallocation and Ownership Transfer}.
\newblock \bibinfo{journal}{\emph{CoRR}}  \bibinfo{volume}{abs/2510.08939}
  (\bibinfo{year}{2025}).
\newblock
\href{https://doi.org/10.48550/ARXIV.2510.08939}{doi:\nolinkurl{10.48550/ARXIV.2510.08939}}


\bibitem[Dennis and Horn(1966)]%
        {DBLP:journals/cacm/DennisH66}
\bibfield{author}{\bibinfo{person}{Jack~B. Dennis} {and} \bibinfo{person}{Earl
  C.~Van Horn}.} \bibinfo{year}{1966}\natexlab{}.
\newblock \showarticletitle{Programming semantics for multiprogrammed
  computations}.
\newblock \bibinfo{journal}{\emph{Commun. {ACM}}} \bibinfo{volume}{9},
  \bibinfo{number}{3} (\bibinfo{year}{1966}), \bibinfo{pages}{143--155}.
\newblock
\href{https://doi.org/10.1145/365230.365252}{doi:\nolinkurl{10.1145/365230.365252}}


\bibitem[Dunfield and Krishnaswami(2022)]%
        {DBLP:journals/csur/DunfieldK21}
\bibfield{author}{\bibinfo{person}{Jana Dunfield} {and} \bibinfo{person}{Neel
  Krishnaswami}.} \bibinfo{year}{2022}\natexlab{}.
\newblock \showarticletitle{Bidirectional Typing}.
\newblock \bibinfo{journal}{\emph{{ACM} Comput. Surv.}} \bibinfo{volume}{54},
  \bibinfo{number}{5} (\bibinfo{year}{2022}), \bibinfo{pages}{98:1--98:38}.
\newblock
\href{https://doi.org/10.1145/3450952}{doi:\nolinkurl{10.1145/3450952}}


\bibitem[Filinski(1994)]%
        {DBLP:conf/popl/Filinski94}
\bibfield{author}{\bibinfo{person}{Andrzej Filinski}.}
  \bibinfo{year}{1994}\natexlab{}.
\newblock \showarticletitle{Representing Monads}. In
  \bibinfo{booktitle}{\emph{{POPL}}}. \bibinfo{publisher}{{ACM} Press},
  \bibinfo{pages}{446--457}.
\newblock
\href{https://doi.org/10.1145/174675.178047}{doi:\nolinkurl{10.1145/174675.178047}}


\bibitem[Filinski(1999)]%
        {DBLP:conf/popl/Filinski99}
\bibfield{author}{\bibinfo{person}{Andrzej Filinski}.}
  \bibinfo{year}{1999}\natexlab{}.
\newblock \showarticletitle{Representing Layered Monads}. In
  \bibinfo{booktitle}{\emph{{POPL}}}. \bibinfo{publisher}{{ACM}},
  \bibinfo{pages}{175--188}.
\newblock
\href{https://doi.org/10.1145/292540.292557}{doi:\nolinkurl{10.1145/292540.292557}}


\bibitem[Fink et~al\mbox{.}(2008)]%
        {DBLP:journals/tosem/FinkYDRG08}
\bibfield{author}{\bibinfo{person}{Stephen~J. Fink}, \bibinfo{person}{Eran
  Yahav}, \bibinfo{person}{Nurit Dor}, \bibinfo{person}{G. Ramalingam}, {and}
  \bibinfo{person}{Emmanuel Geay}.} \bibinfo{year}{2008}\natexlab{}.
\newblock \showarticletitle{Effective typestate verification in the presence of
  aliasing}.
\newblock \bibinfo{journal}{\emph{{ACM} Trans. Softw. Eng. Methodol.}}
  \bibinfo{volume}{17}, \bibinfo{number}{2} (\bibinfo{year}{2008}),
  \bibinfo{pages}{9:1--9:34}.
\newblock
\href{https://doi.org/10.1145/1348250.1348255}{doi:\nolinkurl{10.1145/1348250.1348255}}


\bibitem[Flanagan et~al\mbox{.}(1993)]%
        {DBLP:conf/pldi/FlanaganSDF93}
\bibfield{author}{\bibinfo{person}{Cormac Flanagan}, \bibinfo{person}{Amr
  Sabry}, \bibinfo{person}{Bruce~F. Duba}, {and} \bibinfo{person}{Matthias
  Felleisen}.} \bibinfo{year}{1993}\natexlab{}.
\newblock \showarticletitle{The Essence of Compiling with Continuations}. In
  \bibinfo{booktitle}{\emph{{PLDI}}}. \bibinfo{publisher}{{ACM}},
  \bibinfo{pages}{237--247}.
\newblock
\href{https://doi.org/10.1145/155090.155113}{doi:\nolinkurl{10.1145/155090.155113}}


\bibitem[Garcia et~al\mbox{.}(2014)]%
        {DBLP:journals/toplas/GarciaTWA14}
\bibfield{author}{\bibinfo{person}{Ronald Garcia}, \bibinfo{person}{{\'{E}}ric
  Tanter}, \bibinfo{person}{Roger Wolff}, {and} \bibinfo{person}{Jonathan
  Aldrich}.} \bibinfo{year}{2014}\natexlab{}.
\newblock \showarticletitle{Foundations of Typestate-Oriented Programming}.
\newblock \bibinfo{journal}{\emph{{ACM} Trans. Program. Lang. Syst.}}
  \bibinfo{volume}{36}, \bibinfo{number}{4} (\bibinfo{year}{2014}),
  \bibinfo{pages}{12:1--12:44}.
\newblock
\href{https://doi.org/10.1145/2629609}{doi:\nolinkurl{10.1145/2629609}}


\bibitem[Gay et~al\mbox{.}(2010)]%
        {DBLP:conf/popl/GayVRGC10}
\bibfield{author}{\bibinfo{person}{Simon~J. Gay},
  \bibinfo{person}{Vasco~Thudichum Vasconcelos}, \bibinfo{person}{Ant{\'{o}}nio
  Ravara}, \bibinfo{person}{Nils Gesbert}, {and} \bibinfo{person}{Alexandre~Z.
  Caldeira}.} \bibinfo{year}{2010}\natexlab{}.
\newblock \showarticletitle{Modular session types for distributed
  object-oriented programming}. In \bibinfo{booktitle}{\emph{{POPL}}}.
  \bibinfo{publisher}{{ACM}}, \bibinfo{pages}{299--312}.
\newblock
\href{https://doi.org/10.1145/1706299.1706335}{doi:\nolinkurl{10.1145/1706299.1706335}}


\bibitem[Girard(1987)]%
        {DBLP:journals/tcs/Girard87}
\bibfield{author}{\bibinfo{person}{Jean{-}Yves Girard}.}
  \bibinfo{year}{1987}\natexlab{}.
\newblock \showarticletitle{Linear Logic}.
\newblock \bibinfo{journal}{\emph{Theor. Comput. Sci.}}  \bibinfo{volume}{50}
  (\bibinfo{year}{1987}), \bibinfo{pages}{1--102}.
\newblock
\href{https://doi.org/10.1016/0304-3975(87)90045-4}{doi:\nolinkurl{10.1016/0304-3975(87)90045-4}}


\bibitem[Gordon(2021)]%
        {DBLP:journals/toplas/Gordon21}
\bibfield{author}{\bibinfo{person}{Colin~S. Gordon}.}
  \bibinfo{year}{2021}\natexlab{}.
\newblock \showarticletitle{Polymorphic Iterable Sequential Effect Systems}.
\newblock \bibinfo{journal}{\emph{{ACM} Trans. Program. Lang. Syst.}}
  \bibinfo{volume}{43}, \bibinfo{number}{1} (\bibinfo{year}{2021}),
  \bibinfo{pages}{4:1--4:79}.
\newblock
\href{https://doi.org/10.1145/3450272}{doi:\nolinkurl{10.1145/3450272}}


\bibitem[He et~al\mbox{.}(2026)]%
        {he26arena}
\bibfield{author}{\bibinfo{person}{Siyuan He}, \bibinfo{person}{Songlin Jia},
  \bibinfo{person}{Yuyan Bao}, {and} \bibinfo{person}{Tiark Rompf}.}
  \bibinfo{year}{2026}\natexlab{}.
\newblock \showarticletitle{When {{Lifetimes Liberate}}: {{A Type System}} for
  {{Arenas}} with {{Higher-Order Reachability Tracking}}}.
\newblock \bibinfo{journal}{\emph{Proc. {ACM} Program. Lang.}}
  \bibinfo{volume}{10}, \bibinfo{number}{{OOPSLA1}} (\bibinfo{year}{2026}),
  \bibinfo{pages}{1486--1513}.
\newblock
\href{https://doi.org/10.1145/3798254}{doi:\nolinkurl{10.1145/3798254}}


\bibitem[Henglein et~al\mbox{.}(2005)]%
        {hengleinEffectTypesRegionBased2005}
\bibfield{author}{\bibinfo{person}{Fritz Henglein}, \bibinfo{person}{Henning
  Makholm}, {and} \bibinfo{person}{Henning Niss}.}
  \bibinfo{year}{2005}\natexlab{}.
\newblock \showarticletitle{Effect {{Types}} and {{Region-Based Memory
  Management}}}.
\newblock In \bibinfo{booktitle}{\emph{Advanced Topics in Types and Programming
  Languages}}, \bibfield{editor}{\bibinfo{person}{Benjamin~C. Pierce}} (Ed.).
  \bibinfo{publisher}{MIT Press}, \bibinfo{address}{Cambridge, Mass}.
\newblock
\showISBNx{978-0-262-16228-9}
\showLCCN{QA76.7 .A36 2005}


\bibitem[Honda(1993)]%
        {DBLP:conf/concur/Honda93}
\bibfield{author}{\bibinfo{person}{Kohei Honda}.}
  \bibinfo{year}{1993}\natexlab{}.
\newblock \showarticletitle{Types for Dyadic Interaction}. In
  \bibinfo{booktitle}{\emph{{CONCUR}}} \emph{(\bibinfo{series}{Lecture Notes in
  Computer Science})}. \bibinfo{publisher}{Springer},
  \bibinfo{pages}{509--523}.
\newblock
\href{https://doi.org/10.1007/3-540-57208-2\_35}{doi:\nolinkurl{10.1007/3-540-57208-2\_35}}


\bibitem[Honda et~al\mbox{.}(1998)]%
        {DBLP:conf/esop/HondaVK98}
\bibfield{author}{\bibinfo{person}{Kohei Honda},
  \bibinfo{person}{Vasco~Thudichum Vasconcelos}, {and} \bibinfo{person}{Makoto
  Kubo}.} \bibinfo{year}{1998}\natexlab{}.
\newblock \showarticletitle{Language Primitives and Type Discipline for
  Structured Communication-Based Programming}. In
  \bibinfo{booktitle}{\emph{{ESOP}}} \emph{(\bibinfo{series}{Lecture Notes in
  Computer Science})}. \bibinfo{publisher}{Springer},
  \bibinfo{pages}{122--138}.
\newblock
\href{https://doi.org/10.1007/BFB0053567}{doi:\nolinkurl{10.1007/BFB0053567}}


\bibitem[H{\"{u}}ttel et~al\mbox{.}(2016)]%
        {DBLP:journals/csur/HuttelLVCCDMPRT16}
\bibfield{author}{\bibinfo{person}{Hans H{\"{u}}ttel}, \bibinfo{person}{Ivan
  Lanese}, \bibinfo{person}{Vasco~T. Vasconcelos}, \bibinfo{person}{Lu{\'{\i}}s
  Caires}, \bibinfo{person}{Marco Carbone}, \bibinfo{person}{Pierre{-}Malo
  Deni{\'{e}}lou}, \bibinfo{person}{Dimitris Mostrous}, \bibinfo{person}{Luca
  Padovani}, \bibinfo{person}{Ant{\'{o}}nio Ravara}, \bibinfo{person}{Emilio
  Tuosto}, \bibinfo{person}{Hugo~Torres Vieira}, {and}
  \bibinfo{person}{Gianluigi Zavattaro}.} \bibinfo{year}{2016}\natexlab{}.
\newblock \showarticletitle{Foundations of Session Types and Behavioural
  Contracts}.
\newblock \bibinfo{journal}{\emph{{ACM} Comput. Surv.}} \bibinfo{volume}{49},
  \bibinfo{number}{1} (\bibinfo{year}{2016}), \bibinfo{pages}{3:1--3:36}.
\newblock
\href{https://doi.org/10.1145/2873052}{doi:\nolinkurl{10.1145/2873052}}


\bibitem[Jakobsen et~al\mbox{.}(2021)]%
        {DBLP:conf/ppdp/JakobsenRD21}
\bibfield{author}{\bibinfo{person}{Mathias Jakobsen}, \bibinfo{person}{Alice
  Ravier}, {and} \bibinfo{person}{Ornela Dardha}.}
  \bibinfo{year}{2021}\natexlab{}.
\newblock \showarticletitle{Papaya: Global Typestate Analysis of Aliased
  Objects}. In \bibinfo{booktitle}{\emph{{PPDP}}}. \bibinfo{publisher}{{ACM}},
  \bibinfo{pages}{19:1--19:13}.
\newblock
\href{https://doi.org/10.1145/3479394.3479414}{doi:\nolinkurl{10.1145/3479394.3479414}}


\bibitem[Jespersen et~al\mbox{.}(2015)]%
        {DBLP:conf/icfp/JespersenML15}
\bibfield{author}{\bibinfo{person}{Thomas Bracht~Laumann Jespersen},
  \bibinfo{person}{Philip Munksgaard}, {and} \bibinfo{person}{Ken~Friis
  Larsen}.} \bibinfo{year}{2015}\natexlab{}.
\newblock \showarticletitle{Session types for Rust}. In
  \bibinfo{booktitle}{\emph{WGP@ICFP}}. \bibinfo{publisher}{{ACM}},
  \bibinfo{pages}{13--22}.
\newblock
\href{https://doi.org/10.1145/2808098.2808100}{doi:\nolinkurl{10.1145/2808098.2808100}}


\bibitem[Jia et~al\mbox{.}(2026a)]%
        {artifact}
\bibfield{author}{\bibinfo{person}{Songlin Jia}, \bibinfo{person}{Craig Liu},
  \bibinfo{person}{Siyuan He}, \bibinfo{person}{Haotian Deng},
  \bibinfo{person}{Yuyan Bao}, {and} \bibinfo{person}{Tiark Rompf}.}
  \bibinfo{year}{2026}\natexlab{a}.
\newblock \bibinfo{booktitle}{\emph{Artifact for "Typestate via Revocable
  Capabilities"}}.
\newblock
\href{https://doi.org/10.5281/zenodo.19614655}{doi:\nolinkurl{10.5281/zenodo.19614655}}


\bibitem[Jia et~al\mbox{.}(2026b)]%
        {jia26escape}
\bibfield{author}{\bibinfo{person}{Songlin Jia}, \bibinfo{person}{Guannan Wei},
  \bibinfo{person}{Siyuan He}, \bibinfo{person}{Yuyan Bao}, {and}
  \bibinfo{person}{Tiark Rompf}.} \bibinfo{year}{2026}\natexlab{b}.
\newblock \showarticletitle{Escape with Your Self: Sound and Expressive
  Bidirectional Typing with Avoidance for Reachability Types}.
\newblock \bibinfo{journal}{\emph{Proc. {ACM} Program. Lang.}}
  \bibinfo{volume}{10}, \bibinfo{number}{{PLDI}} (\bibinfo{year}{2026}).
\newblock
\href{https://doi.org/10.1145/3808335}{doi:\nolinkurl{10.1145/3808335}}


\bibitem[Kiselyov and Ishii(2015)]%
        {DBLP:conf/haskell/KiselyovI15}
\bibfield{author}{\bibinfo{person}{Oleg Kiselyov} {and} \bibinfo{person}{Hiromi
  Ishii}.} \bibinfo{year}{2015}\natexlab{}.
\newblock \showarticletitle{Freer monads, more extensible effects}. In
  \bibinfo{booktitle}{\emph{Haskell}}. \bibinfo{publisher}{{ACM}},
  \bibinfo{pages}{94--105}.
\newblock
\href{https://doi.org/10.1145/2804302.2804319}{doi:\nolinkurl{10.1145/2804302.2804319}}


\bibitem[Kiselyov et~al\mbox{.}(2004)]%
        {DBLP:conf/haskell/KiselyovLS04}
\bibfield{author}{\bibinfo{person}{Oleg Kiselyov}, \bibinfo{person}{Ralf
  L{\"{a}}mmel}, {and} \bibinfo{person}{Keean Schupke}.}
  \bibinfo{year}{2004}\natexlab{}.
\newblock \showarticletitle{Strongly typed heterogeneous collections}. In
  \bibinfo{booktitle}{\emph{Proceedings of the {ACM} {SIGPLAN} Workshop on
  Haskell, Haskell 2004, Snowbird, UT, USA, September 22-22, 2004}}.
  \bibinfo{publisher}{{ACM}}, \bibinfo{pages}{96--107}.
\newblock
\href{https://doi.org/10.1145/1017472.1017488}{doi:\nolinkurl{10.1145/1017472.1017488}}


\bibitem[Kokke(2019)]%
        {DBLP:journals/corr/abs-1909-05970}
\bibfield{author}{\bibinfo{person}{Wen Kokke}.}
  \bibinfo{year}{2019}\natexlab{}.
\newblock \showarticletitle{Rusty Variation: Deadlock-free Sessions with
  Failure in Rust}. In \bibinfo{booktitle}{\emph{{ICE}}}
  \emph{(\bibinfo{series}{{EPTCS}})}. \bibinfo{pages}{48--60}.
\newblock
\href{https://doi.org/10.4204/EPTCS.304.4}{doi:\nolinkurl{10.4204/EPTCS.304.4}}


\bibitem[Liang et~al\mbox{.}(1995)]%
        {DBLP:conf/popl/LiangHJ95}
\bibfield{author}{\bibinfo{person}{Sheng Liang}, \bibinfo{person}{Paul Hudak},
  {and} \bibinfo{person}{Mark~P. Jones}.} \bibinfo{year}{1995}\natexlab{}.
\newblock \showarticletitle{Monad Transformers and Modular Interpreters}. In
  \bibinfo{booktitle}{\emph{{POPL}}}. \bibinfo{publisher}{{ACM} Press},
  \bibinfo{pages}{333--343}.
\newblock
\href{https://doi.org/10.1145/199448.199528}{doi:\nolinkurl{10.1145/199448.199528}}


\bibitem[Lindley et~al\mbox{.}(2017)]%
        {DBLP:conf/popl/LindleyMM17}
\bibfield{author}{\bibinfo{person}{Sam Lindley}, \bibinfo{person}{Conor
  McBride}, {and} \bibinfo{person}{Craig McLaughlin}.}
  \bibinfo{year}{2017}\natexlab{}.
\newblock \showarticletitle{Do be do be do}. In
  \bibinfo{booktitle}{\emph{{POPL}}}. \bibinfo{publisher}{{ACM}},
  \bibinfo{pages}{500--514}.
\newblock
\href{https://doi.org/10.1145/3009837.3009897}{doi:\nolinkurl{10.1145/3009837.3009897}}


\bibitem[Lucassen and Gifford(1988)]%
        {DBLP:conf/popl/LucassenG88}
\bibfield{author}{\bibinfo{person}{John~M. Lucassen} {and}
  \bibinfo{person}{David~K. Gifford}.} \bibinfo{year}{1988}\natexlab{}.
\newblock \showarticletitle{Polymorphic Effect Systems}. In
  \bibinfo{booktitle}{\emph{{POPL}}}. \bibinfo{publisher}{{ACM} Press},
  \bibinfo{pages}{47--57}.
\newblock
\href{https://doi.org/10.1145/73560.73564}{doi:\nolinkurl{10.1145/73560.73564}}


\bibitem[Marino and Millstein(2009)]%
        {DBLP:conf/tldi/MarinoM09}
\bibfield{author}{\bibinfo{person}{Daniel Marino} {and}
  \bibinfo{person}{Todd~D. Millstein}.} \bibinfo{year}{2009}\natexlab{}.
\newblock \showarticletitle{A generic type-and-effect system}. In
  \bibinfo{booktitle}{\emph{{TLDI}}}. \bibinfo{publisher}{{ACM}},
  \bibinfo{pages}{39--50}.
\newblock
\href{https://doi.org/10.1145/1481861.1481868}{doi:\nolinkurl{10.1145/1481861.1481868}}


\bibitem[Matsakis and II(2014)]%
        {DBLP:conf/sigada/MatsakisK14}
\bibfield{author}{\bibinfo{person}{Nicholas~D. Matsakis} {and}
  \bibinfo{person}{Felix S.~Klock II}.} \bibinfo{year}{2014}\natexlab{}.
\newblock \showarticletitle{The rust language}. In
  \bibinfo{booktitle}{\emph{{HILT}}}. \bibinfo{publisher}{{ACM}},
  \bibinfo{pages}{103--104}.
\newblock
\href{https://doi.org/10.1145/2663171.2663188}{doi:\nolinkurl{10.1145/2663171.2663188}}


\bibitem[Miller and Shapiro(2003)]%
        {DBLP:conf/asian/MillerS03}
\bibfield{author}{\bibinfo{person}{Mark~S. Miller} {and}
  \bibinfo{person}{Jonathan~S. Shapiro}.} \bibinfo{year}{2003}\natexlab{}.
\newblock \showarticletitle{Paradigm Regained: Abstraction Mechanisms for
  Access Control}. In \bibinfo{booktitle}{\emph{{ASIAN}}}
  \emph{(\bibinfo{series}{Lecture Notes in Computer Science},
  Vol.~\bibinfo{volume}{2896})}. \bibinfo{publisher}{Springer},
  \bibinfo{pages}{224--242}.
\newblock
\href{https://doi.org/10.1007/978-3-540-40965-6\_15}{doi:\nolinkurl{10.1007/978-3-540-40965-6\_15}}


\bibitem[Naeem and Lhot{\'{a}}k(2008)]%
        {DBLP:conf/oopsla/NaeemL08}
\bibfield{author}{\bibinfo{person}{Nomair~A. Naeem} {and}
  \bibinfo{person}{Ondrej Lhot{\'{a}}k}.} \bibinfo{year}{2008}\natexlab{}.
\newblock \showarticletitle{Typestate-like analysis of multiple interacting
  objects}. In \bibinfo{booktitle}{\emph{{OOPSLA}}}.
  \bibinfo{publisher}{{ACM}}, \bibinfo{pages}{347--366}.
\newblock
\href{https://doi.org/10.1145/1449764.1449792}{doi:\nolinkurl{10.1145/1449764.1449792}}


\bibitem[Odersky et~al\mbox{.}(2023)]%
        {ScalaCC}
\bibfield{author}{\bibinfo{person}{Martin Odersky} {et~al\mbox{.}}}
  \bibinfo{year}{2023}\natexlab{}.
\newblock \bibinfo{booktitle}{\emph{{Scala 3 Reference - Capture Checking}}}.
\newblock
\urldef\tempurl%
\url{https://docs.scala-lang.org/scala3/reference/experimental/cc.html}
\showURL{%
\tempurl}


\bibitem[Odersky et~al\mbox{.}(2025)]%
        {SeparationChecking}
\bibfield{author}{\bibinfo{person}{Martin Odersky} {et~al\mbox{.}}}
  \bibinfo{year}{2025}\natexlab{}.
\newblock \bibinfo{booktitle}{\emph{Separation {{Checking}}}}.
\newblock
\urldef\tempurl%
\url{https://nightly.scala-lang.org/docs/reference/experimental/capture-checking/separation-checking.html}
\showURL{%
\tempurl}


\bibitem[Odersky et~al\mbox{.}(2018)]%
        {DBLP:journals/pacmpl/OderskyBLBMS18}
\bibfield{author}{\bibinfo{person}{Martin Odersky}, \bibinfo{person}{Olivier
  Blanvillain}, \bibinfo{person}{Fengyun Liu}, \bibinfo{person}{Aggelos
  Biboudis}, \bibinfo{person}{Heather Miller}, {and} \bibinfo{person}{Sandro
  Stucki}.} \bibinfo{year}{2018}\natexlab{}.
\newblock \showarticletitle{Simplicitly: foundations and applications of
  implicit function types}.
\newblock \bibinfo{journal}{\emph{Proc. {ACM} Program. Lang.}}
  \bibinfo{volume}{2}, \bibinfo{number}{{POPL}} (\bibinfo{year}{2018}),
  \bibinfo{pages}{42:1--42:29}.
\newblock
\href{https://doi.org/10.1145/3158130}{doi:\nolinkurl{10.1145/3158130}}


\bibitem[Odersky et~al\mbox{.}(2021)]%
        {DBLP:conf/scala/OderskyBBLL21}
\bibfield{author}{\bibinfo{person}{Martin Odersky}, \bibinfo{person}{Aleksander
  Boruch{-}Gruszecki}, \bibinfo{person}{Jonathan~Immanuel Brachth{\"{a}}user},
  \bibinfo{person}{Edward Lee}, {and} \bibinfo{person}{Ondrej Lhot{\'{a}}k}.}
  \bibinfo{year}{2021}\natexlab{}.
\newblock \showarticletitle{Safer exceptions for Scala}. In
  \bibinfo{booktitle}{\emph{SCALA/SPLASH}}. \bibinfo{publisher}{{ACM}},
  \bibinfo{pages}{1--11}.
\newblock
\href{https://doi.org/10.1145/3486610.3486893}{doi:\nolinkurl{10.1145/3486610.3486893}}


\bibitem[Odersky et~al\mbox{.}(2001)]%
        {DBLP:conf/popl/OderskyZZ01}
\bibfield{author}{\bibinfo{person}{Martin Odersky}, \bibinfo{person}{Christoph
  Zenger}, {and} \bibinfo{person}{Matthias Zenger}.}
  \bibinfo{year}{2001}\natexlab{}.
\newblock \showarticletitle{Colored local type inference}. In
  \bibinfo{booktitle}{\emph{{POPL}}}. \bibinfo{publisher}{{ACM}},
  \bibinfo{pages}{41--53}.
\newblock
\href{https://doi.org/10.1145/360204.360207}{doi:\nolinkurl{10.1145/360204.360207}}


\bibitem[Odersky and Zenger(2005)]%
        {DBLP:conf/oopsla/OderskyZ05}
\bibfield{author}{\bibinfo{person}{Martin Odersky} {and}
  \bibinfo{person}{Matthias Zenger}.} \bibinfo{year}{2005}\natexlab{}.
\newblock \showarticletitle{Scalable component abstractions}. In
  \bibinfo{booktitle}{\emph{{OOPSLA}}}. \bibinfo{publisher}{{ACM}},
  \bibinfo{pages}{41--57}.
\newblock
\href{https://doi.org/10.1145/1094811.1094815}{doi:\nolinkurl{10.1145/1094811.1094815}}


\bibitem[Osvald et~al\mbox{.}(2016)]%
        {DBLP:conf/oopsla/OsvaldEWAR16}
\bibfield{author}{\bibinfo{person}{Leo Osvald},
  \bibinfo{person}{Gr{\'{e}}gory~M. Essertel}, \bibinfo{person}{Xilun Wu},
  \bibinfo{person}{Lilliam I.~Gonz{\'{a}}lez Alay{\'{o}}n}, {and}
  \bibinfo{person}{Tiark Rompf}.} \bibinfo{year}{2016}\natexlab{}.
\newblock \showarticletitle{Gentrification gone too far? affordable 2nd-class
  values for fun and (co-)effect}. In \bibinfo{booktitle}{\emph{{OOPSLA}}}.
  \bibinfo{publisher}{{ACM}}, \bibinfo{pages}{234--251}.
\newblock
\href{https://doi.org/10.1145/2983990.2984009}{doi:\nolinkurl{10.1145/2983990.2984009}}


\bibitem[Pucella and Tov(2008)]%
        {DBLP:conf/haskell/PucellaT08}
\bibfield{author}{\bibinfo{person}{Riccardo Pucella} {and}
  \bibinfo{person}{Jesse~A. Tov}.} \bibinfo{year}{2008}\natexlab{}.
\newblock \showarticletitle{Haskell session types with (almost) no class}. In
  \bibinfo{booktitle}{\emph{Haskell}}. \bibinfo{publisher}{{ACM}},
  \bibinfo{pages}{25--36}.
\newblock
\href{https://doi.org/10.1145/1411286.1411290}{doi:\nolinkurl{10.1145/1411286.1411290}}


\bibitem[Rompf and Amin(2016)]%
        {DBLP:conf/oopsla/RompfA16}
\bibfield{author}{\bibinfo{person}{Tiark Rompf} {and} \bibinfo{person}{Nada
  Amin}.} \bibinfo{year}{2016}\natexlab{}.
\newblock \showarticletitle{Type soundness for dependent object types {(DOT)}}.
  In \bibinfo{booktitle}{\emph{{OOPSLA}}}. \bibinfo{publisher}{{ACM}},
  \bibinfo{pages}{624--641}.
\newblock
\href{https://doi.org/10.1145/2983990.2984008}{doi:\nolinkurl{10.1145/2983990.2984008}}


\bibitem[Rompf et~al\mbox{.}(2009)]%
        {DBLP:conf/icfp/RompfMO09}
\bibfield{author}{\bibinfo{person}{Tiark Rompf}, \bibinfo{person}{Ingo Maier},
  {and} \bibinfo{person}{Martin Odersky}.} \bibinfo{year}{2009}\natexlab{}.
\newblock \showarticletitle{Implementing first-class polymorphic delimited
  continuations by a type-directed selective CPS-transform}. In
  \bibinfo{booktitle}{\emph{{ICFP}}}. \bibinfo{publisher}{{ACM}},
  \bibinfo{pages}{317--328}.
\newblock
\href{https://doi.org/10.1145/1596550.1596596}{doi:\nolinkurl{10.1145/1596550.1596596}}


\bibitem[Rytz et~al\mbox{.}(2012)]%
        {DBLP:conf/ecoop/RytzOH12}
\bibfield{author}{\bibinfo{person}{Lukas Rytz}, \bibinfo{person}{Martin
  Odersky}, {and} \bibinfo{person}{Philipp Haller}.}
  \bibinfo{year}{2012}\natexlab{}.
\newblock \showarticletitle{Lightweight Polymorphic Effects}. In
  \bibinfo{booktitle}{\emph{{ECOOP}}} \emph{(\bibinfo{series}{Lecture Notes in
  Computer Science}, Vol.~\bibinfo{volume}{7313})}.
  \bibinfo{publisher}{Springer}, \bibinfo{pages}{258--282}.
\newblock
\href{https://doi.org/10.1007/978-3-642-31057-7\_13}{doi:\nolinkurl{10.1007/978-3-642-31057-7\_13}}


\bibitem[Saffrich et~al\mbox{.}(2024)]%
        {DBLP:journals/pacmpl/Saffrich0024}
\bibfield{author}{\bibinfo{person}{Hannes Saffrich}, \bibinfo{person}{Yuki
  Nishida}, {and} \bibinfo{person}{Peter Thiemann}.}
  \bibinfo{year}{2024}\natexlab{}.
\newblock \showarticletitle{Law and Order for Typestate with Borrowing}.
\newblock \bibinfo{journal}{\emph{Proc. {ACM} Program. Lang.}}
  \bibinfo{volume}{8}, \bibinfo{number}{{OOPSLA2}} (\bibinfo{year}{2024}),
  \bibinfo{pages}{1475--1503}.
\newblock
\href{https://doi.org/10.1145/3689763}{doi:\nolinkurl{10.1145/3689763}}


\bibitem[Saffrich et~al\mbox{.}(2025)]%
        {DBLP:journals/pacmpl/SaffrichS0V25}
\bibfield{author}{\bibinfo{person}{Hannes Saffrich}, \bibinfo{person}{Janek
  Spaderna}, \bibinfo{person}{Peter Thiemann}, {and} \bibinfo{person}{Vasco~T.
  Vasconcelos}.} \bibinfo{year}{2025}\natexlab{}.
\newblock \showarticletitle{Borrowing from Session Types}.
\newblock \bibinfo{journal}{\emph{Proc. {ACM} Program. Lang.}}
  \bibinfo{volume}{9}, \bibinfo{number}{{OOPSLA2}} (\bibinfo{year}{2025}),
  \bibinfo{pages}{3426--3453}.
\newblock
\href{https://doi.org/10.1145/3763173}{doi:\nolinkurl{10.1145/3763173}}


\bibitem[Saffrich and Thiemann(2022)]%
        {DBLP:journals/lmcs/SaffrichT22}
\bibfield{author}{\bibinfo{person}{Hannes Saffrich} {and}
  \bibinfo{person}{Peter Thiemann}.} \bibinfo{year}{2022}\natexlab{}.
\newblock \showarticletitle{Relating Functional and Imperative Session Types}.
\newblock \bibinfo{journal}{\emph{Log. Methods Comput. Sci.}}
  \bibinfo{volume}{18}, \bibinfo{number}{3} (\bibinfo{year}{2022}).
\newblock
\href{https://doi.org/10.46298/LMCS-18(3:33)2022}{doi:\nolinkurl{10.46298/LMCS-18(3:33)2022}}


\bibitem[Saffrich and Thiemann(2023)]%
        {DBLP:conf/ppdp/Saffrich023}
\bibfield{author}{\bibinfo{person}{Hannes Saffrich} {and}
  \bibinfo{person}{Peter Thiemann}.} \bibinfo{year}{2023}\natexlab{}.
\newblock \showarticletitle{Polymorphic Typestate for Session Types}. In
  \bibinfo{booktitle}{\emph{{PPDP}}}. \bibinfo{publisher}{{ACM}},
  \bibinfo{pages}{12:1--12:15}.
\newblock
\href{https://doi.org/10.1145/3610612.3610624}{doi:\nolinkurl{10.1145/3610612.3610624}}


\bibitem[Scalas and Yoshida(2016)]%
        {DBLP:conf/ecoop/ScalasY16}
\bibfield{author}{\bibinfo{person}{Alceste Scalas} {and}
  \bibinfo{person}{Nobuko Yoshida}.} \bibinfo{year}{2016}\natexlab{}.
\newblock \showarticletitle{Lightweight Session Programming in Scala}. In
  \bibinfo{booktitle}{\emph{{ECOOP}}} \emph{(\bibinfo{series}{LIPIcs})}.
  \bibinfo{publisher}{Schloss Dagstuhl - Leibniz-Zentrum f{\"{u}}r Informatik},
  \bibinfo{pages}{21:1--21:28}.
\newblock
\href{https://doi.org/10.4230/LIPICS.ECOOP.2016.21}{doi:\nolinkurl{10.4230/LIPICS.ECOOP.2016.21}}


\bibitem[Spiwack et~al\mbox{.}(2022)]%
        {DBLP:journals/pacmpl/SpiwackKBWE22}
\bibfield{author}{\bibinfo{person}{Arnaud Spiwack}, \bibinfo{person}{Csongor
  Kiss}, \bibinfo{person}{Jean{-}Philippe Bernardy}, \bibinfo{person}{Nicolas
  Wu}, {and} \bibinfo{person}{Richard~A. Eisenberg}.}
  \bibinfo{year}{2022}\natexlab{}.
\newblock \showarticletitle{Linearly qualified types: generic inference for
  capabilities and uniqueness}.
\newblock \bibinfo{journal}{\emph{Proc. {ACM} Program. Lang.}}
  \bibinfo{volume}{6}, \bibinfo{number}{{ICFP}} (\bibinfo{year}{2022}),
  \bibinfo{pages}{137--164}.
\newblock
\href{https://doi.org/10.1145/3547626}{doi:\nolinkurl{10.1145/3547626}}


\bibitem[Strom and Yemini(1986)]%
        {DBLP:journals/tse/StromY86}
\bibfield{author}{\bibinfo{person}{Robert~E. Strom} {and}
  \bibinfo{person}{Shaula Yemini}.} \bibinfo{year}{1986}\natexlab{}.
\newblock \showarticletitle{Typestate: {A} Programming Language Concept for
  Enhancing Software Reliability}.
\newblock \bibinfo{journal}{\emph{{IEEE} Trans. Software Eng.}}
  \bibinfo{volume}{12}, \bibinfo{number}{1} (\bibinfo{year}{1986}),
  \bibinfo{pages}{157--171}.
\newblock
\href{https://doi.org/10.1109/TSE.1986.6312929}{doi:\nolinkurl{10.1109/TSE.1986.6312929}}


\bibitem[Swierstra(2008)]%
        {DBLP:journals/jfp/Swierstra08}
\bibfield{author}{\bibinfo{person}{Wouter Swierstra}.}
  \bibinfo{year}{2008}\natexlab{}.
\newblock \showarticletitle{Data types {\`{a}} la carte}.
\newblock \bibinfo{journal}{\emph{J. Funct. Program.}} \bibinfo{volume}{18},
  \bibinfo{number}{4} (\bibinfo{year}{2008}), \bibinfo{pages}{423--436}.
\newblock
\href{https://doi.org/10.1017/S0956796808006758}{doi:\nolinkurl{10.1017/S0956796808006758}}


\bibitem[Takeuchi et~al\mbox{.}(1994)]%
        {DBLP:conf/parle/TakeuchiHK94}
\bibfield{author}{\bibinfo{person}{Kaku Takeuchi}, \bibinfo{person}{Kohei
  Honda}, {and} \bibinfo{person}{Makoto Kubo}.}
  \bibinfo{year}{1994}\natexlab{}.
\newblock \showarticletitle{An Interaction-based Language and its Typing
  System}. In \bibinfo{booktitle}{\emph{{PARLE}}}
  \emph{(\bibinfo{series}{Lecture Notes in Computer Science})}.
  \bibinfo{publisher}{Springer}, \bibinfo{pages}{398--413}.
\newblock
\href{https://doi.org/10.1007/3-540-58184-7\_118}{doi:\nolinkurl{10.1007/3-540-58184-7\_118}}


\bibitem[Tang et~al\mbox{.}(2025)]%
        {DBLP:journals/pacmpl/TangWDHLL25}
\bibfield{author}{\bibinfo{person}{Wenhao Tang}, \bibinfo{person}{Leo White},
  \bibinfo{person}{Stephen Dolan}, \bibinfo{person}{Daniel Hillerstr{\"{o}}m},
  \bibinfo{person}{Sam Lindley}, {and} \bibinfo{person}{Anton Lorenzen}.}
  \bibinfo{year}{2025}\natexlab{}.
\newblock \showarticletitle{Modal Effect Types}.
\newblock \bibinfo{journal}{\emph{Proc. {ACM} Program. Lang.}}
  \bibinfo{volume}{9}, \bibinfo{number}{{OOPSLA1}} (\bibinfo{year}{2025}),
  \bibinfo{pages}{1130--1157}.
\newblock
\href{https://doi.org/10.1145/3720476}{doi:\nolinkurl{10.1145/3720476}}


\bibitem[Toro and Tanter(2015)]%
        {DBLP:conf/oopsla/ToroT15}
\bibfield{author}{\bibinfo{person}{Mat{\'{\i}}as Toro} {and}
  \bibinfo{person}{{\'{E}}ric Tanter}.} \bibinfo{year}{2015}\natexlab{}.
\newblock \showarticletitle{Customizable gradual polymorphic effects for
  Scala}. In \bibinfo{booktitle}{\emph{{OOPSLA}}}. \bibinfo{publisher}{{ACM}},
  \bibinfo{pages}{935--953}.
\newblock
\href{https://doi.org/10.1145/2814270.2814315}{doi:\nolinkurl{10.1145/2814270.2814315}}


\bibitem[Vasconcelos et~al\mbox{.}(2006)]%
        {DBLP:journals/tcs/VasconcelosGR06}
\bibfield{author}{\bibinfo{person}{Vasco~Thudichum Vasconcelos},
  \bibinfo{person}{Simon~J. Gay}, {and} \bibinfo{person}{Ant{\'{o}}nio
  Ravara}.} \bibinfo{year}{2006}\natexlab{}.
\newblock \showarticletitle{Type checking a multithreaded functional language
  with session types}.
\newblock \bibinfo{journal}{\emph{Theor. Comput. Sci.}} \bibinfo{volume}{368},
  \bibinfo{number}{1-2} (\bibinfo{year}{2006}), \bibinfo{pages}{64--87}.
\newblock
\href{https://doi.org/10.1016/J.TCS.2006.06.028}{doi:\nolinkurl{10.1016/J.TCS.2006.06.028}}


\bibitem[Wadler(1990)]%
        {DBLP:conf/ifip2/Wadler90}
\bibfield{author}{\bibinfo{person}{Philip Wadler}.}
  \bibinfo{year}{1990}\natexlab{}.
\newblock \showarticletitle{Linear Types can Change the World!}. In
  \bibinfo{booktitle}{\emph{Programming Concepts and Methods}}.
  \bibinfo{publisher}{North-Holland}, \bibinfo{pages}{561}.
\newblock


\bibitem[Wadler(1992)]%
        {DBLP:conf/popl/Wadler92}
\bibfield{author}{\bibinfo{person}{Philip Wadler}.}
  \bibinfo{year}{1992}\natexlab{}.
\newblock \showarticletitle{The Essence of Functional Programming}. In
  \bibinfo{booktitle}{\emph{{POPL}}}. \bibinfo{publisher}{{ACM} Press},
  \bibinfo{pages}{1--14}.
\newblock
\href{https://doi.org/10.1145/143165.143169}{doi:\nolinkurl{10.1145/143165.143169}}


\bibitem[Wadler and Thiemann(2003)]%
        {DBLP:journals/tocl/WadlerT03}
\bibfield{author}{\bibinfo{person}{Philip Wadler} {and} \bibinfo{person}{Peter
  Thiemann}.} \bibinfo{year}{2003}\natexlab{}.
\newblock \showarticletitle{The marriage of effects and monads}.
\newblock \bibinfo{journal}{\emph{{ACM} Trans. Comput. Log.}}
  \bibinfo{volume}{4}, \bibinfo{number}{1} (\bibinfo{year}{2003}),
  \bibinfo{pages}{1--32}.
\newblock
\href{https://doi.org/10.1145/601775.601776}{doi:\nolinkurl{10.1145/601775.601776}}


\bibitem[Wei et~al\mbox{.}(2024)]%
        {DBLP:journals/pacmpl/WeiBJBR24}
\bibfield{author}{\bibinfo{person}{Guannan Wei}, \bibinfo{person}{Oliver
  Bracevac}, \bibinfo{person}{Songlin Jia}, \bibinfo{person}{Yuyan Bao}, {and}
  \bibinfo{person}{Tiark Rompf}.} \bibinfo{year}{2024}\natexlab{}.
\newblock \showarticletitle{Polymorphic Reachability Types: Tracking Freshness,
  Aliasing, and Separation in Higher-Order Generic Programs}.
\newblock \bibinfo{journal}{\emph{Proc. {ACM} Program. Lang.}}
  \bibinfo{volume}{8}, \bibinfo{number}{{POPL}} (\bibinfo{year}{2024}),
  \bibinfo{pages}{393--424}.
\newblock
\href{https://doi.org/10.1145/3632856}{doi:\nolinkurl{10.1145/3632856}}


\bibitem[Xhebraj et~al\mbox{.}(2022)]%
        {DBLP:conf/ecoop/XhebrajB0R22}
\bibfield{author}{\bibinfo{person}{Anxhelo Xhebraj}, \bibinfo{person}{Oliver
  Bracevac}, \bibinfo{person}{Guannan Wei}, {and} \bibinfo{person}{Tiark
  Rompf}.} \bibinfo{year}{2022}\natexlab{}.
\newblock \showarticletitle{What If We Don't Pop the Stack? The Return of
  2nd-Class Values}. In \bibinfo{booktitle}{\emph{{ECOOP}}}
  \emph{(\bibinfo{series}{LIPIcs}, Vol.~\bibinfo{volume}{222})}.
  \bibinfo{publisher}{Schloss Dagstuhl - Leibniz-Zentrum f{\"{u}}r Informatik},
  \bibinfo{pages}{15:1--15:29}.
\newblock
\href{https://doi.org/10.4230/LIPICS.ECOOP.2022.15}{doi:\nolinkurl{10.4230/LIPICS.ECOOP.2022.15}}


\bibitem[Xu et~al\mbox{.}(2024)]%
        {DBLP:journals/pacmpl/XuBO24}
\bibfield{author}{\bibinfo{person}{Yichen Xu}, \bibinfo{person}{Aleksander
  Boruch{-}Gruszecki}, {and} \bibinfo{person}{Martin Odersky}.}
  \bibinfo{year}{2024}\natexlab{}.
\newblock \showarticletitle{Degrees of Separation: {A} Flexible Type System for
  Safe Concurrency}.
\newblock \bibinfo{journal}{\emph{Proc. {ACM} Program. Lang.}}
  \bibinfo{volume}{8}, \bibinfo{number}{{OOPSLA1}} (\bibinfo{year}{2024}),
  \bibinfo{pages}{1181--1207}.
\newblock
\href{https://doi.org/10.1145/3649853}{doi:\nolinkurl{10.1145/3649853}}


\bibitem[Xu et~al\mbox{.}(2025)]%
        {DBLP:journals/pacmpl/XuBPO25}
\bibfield{author}{\bibinfo{person}{Yichen Xu}, \bibinfo{person}{Oliver
  Bracevac}, \bibinfo{person}{Cao~Nguyen Pham}, {and} \bibinfo{person}{Martin
  Odersky}.} \bibinfo{year}{2025}\natexlab{}.
\newblock \showarticletitle{What's in the Box: Ergonomic and Expressive Capture
  Tracking over Generic Data Structures}.
\newblock \bibinfo{journal}{\emph{Proc. {ACM} Program. Lang.}}
  \bibinfo{volume}{9}, \bibinfo{number}{{OOPSLA2}} (\bibinfo{year}{2025}),
  \bibinfo{pages}{1726--1753}.
\newblock
\href{https://doi.org/10.1145/3763112}{doi:\nolinkurl{10.1145/3763112}}


\bibitem[Xu and Odersky(2023)]%
        {DBLP:journals/corr/abs-2306-06496}
\bibfield{author}{\bibinfo{person}{Yichen Xu} {and} \bibinfo{person}{Martin
  Odersky}.} \bibinfo{year}{2023}\natexlab{}.
\newblock \showarticletitle{Formalizing Box Inference for Capture Calculus}.
\newblock \bibinfo{journal}{\emph{CoRR}}  \bibinfo{volume}{abs/2306.06496}
  (\bibinfo{year}{2023}).
\newblock
\href{https://doi.org/10.48550/ARXIV.2306.06496}{doi:\nolinkurl{10.48550/ARXIV.2306.06496}}


\end{thebibliography}

\end{document}